\documentclass[twocolumn,floatfix,showpacs,prd,aps,tightenlines,superscriptaddress]{revtex4-1}
\usepackage{amsmath}
\usepackage{graphicx}
\usepackage{psfrag}
\usepackage[dvipsnames]{xcolor}
\usepackage{dcolumn}
\usepackage{bm}
\usepackage[mathscr]{eucal}
\usepackage{mathrsfs}
\usepackage{xspace}
\usepackage{csquotes}
\usepackage{tabularx}
\newcolumntype{Z}{>{\centering\let\newline\\\arraybackslash\hspace{0pt}}X}
\usepackage{siunitx}
\usepackage[caption=false]{subfig}
\usepackage{tikz}
\usetikzlibrary{arrows,calc,decorations.markings,math,arrows.meta}
\usepackage{comment}

\usepackage[colorlinks=true]{hyperref}
\hypersetup{
  citecolor  = blue
}

\def\apj{Astrophys. J.}  
\def\aap{Astron. Astrophys.}
\def\mnras{Mon. Not. R. Astron. Soc.}  
\def\prd{Phys. Rev. D}
\def\prl{Phys. Rev. Lett.}  

\def\apjl{Astrophys. J. Letters}
\def\apjs{Astrophys. J. Supplement Series}
\def\pasj{PASJ}

\def\ET{{\sc Einstein Toolkit}\xspace}
\def\SENR{{\sc SENR/NRPy+}\xspace}

\def\msun{\rm M_{\odot}}
\def\kms{\rm km \, s^{-1}}

\def\simlt{\mathrel{\rlap{\lower 3pt\hbox{$\sim$}}\raise 2.0pt\hbox{$<$}}}
\def\simgt{\mathrel{\rlap{\lower 3pt\hbox{$\sim$}} \raise 2.0pt\hbox{$>$}}}
\def\lsim{\mathrel{\rlap{\lower 3pt\hbox{$\sim$}}\raise 2.0pt\hbox{$<$}}}
\def\gsim{\mathrel{\rlap{\lower 3pt\hbox{$\sim$}} \raise 2.0pt\hbox{$>$}}}

\def\msunpc3{\msun~{\rm {pc^{-3}}}}
\newcommand{\be}{\begin{equation}}
\newcommand{\ee}{\end{equation}}

\def\kms{{\rm\,km\,s^{-1}}}

\newcommand{\bea}{\begin{eqnarray}}
\newcommand{\eea}{\end{eqnarray}}
\newcommand{\beq}{\begin{equation}}
\newcommand{\eeq}{\end{equation}}

\newcommand{\SphericalNR}{{\sc SphericalNR}\xspace}

\begin{document}

\def\fun#1#2{\lower3.6pt\vbox{\baselineskip0pt\lineskip.9pt
  \ialign{$\mathsurround=0pt#1\hfil##\hfil$\crcr#2\crcr\sim\crcr}}}
\def\lap{\mathrel{\mathpalette\fun <}}
\def\gap{\mathrel{\mathpalette\fun >}}
\def\kms{{\rm km\ s}^{-1}}
\def\vk{V_{\rm recoil}}
\def\araa{Annual Reviews of Astronomy \& Astrophysics}

\title{Ameliorating the Courant-Friedrichs-Lewy condition in spherical coordinates: A double 
FFT filter method for general relativistic MHD in dynamical spacetimes}

\author{Liwei Ji}
\email{ljsma@rit.edu}
\affiliation{Center for Computational Relativity and Gravitation, and School of Mathematical Sciences, Rochester Institute of Technology, 85 Lomb Memorial Drive, Rochester, New York 14623, USA}
\author{Vassilios Mewes}
\affiliation{National Center for Computational Sciences, Oak Ridge National Laboratory, P.O. Box 2008, Oak Ridge, Tennessee 37831-6164, USA}
\author{Yosef Zlochower}
\affiliation{Center for Computational Relativity and Gravitation, and School of Mathematical Sciences, Rochester Institute of Technology, 85 Lomb Memorial Drive, Rochester, New York 14623, USA}
\author{Lorenzo Ennoggi}
\affiliation{Center for Computational Relativity and Gravitation, and School of Mathematical Sciences, Rochester Institute of Technology, 85 Lomb Memorial Drive, Rochester, New York 14623, USA}
\author{Federico G. Lopez Armengol}
\affiliation{NPCx, 23150 Fashion Drive, Suite 238 Estero, Florida 33928, USA}
\affiliation{Center for Computational Relativity and Gravitation, and School of Mathematical Sciences, Rochester Institute of Technology, 85 Lomb Memorial Drive, Rochester, New York 14623, USA}
\author{Manuela Campanelli}
\affiliation{Center for Computational Relativity and Gravitation, and School of Mathematical Sciences, Rochester Institute of Technology, 85 Lomb Memorial Drive, Rochester, New York 14623, USA}
\author{Federico Cipolletta}
\affiliation{Barcelona Supercomputing Center (BSC)}
\author{Zachariah B. Etienne}
\affiliation{Department of Physics, University of Idaho, Moscow, Idaho 83843, USA}
\affiliation{Department of Physics and Astronomy, West Virginia University, Morgantown, West Virginia 26506, USA}
\affiliation{Center for Gravitational Waves and Cosmology, West Virginia University, Chestnut Ridge Research Building, Morgantown, WV 26505, USA}

\begin{abstract}
 Numerical simulations of merging compact objects and their remnants form the theoretical foundation for gravitational wave and multimessenger astronomy. While Cartesian-coordinate-based adaptive mesh refinement is commonly used for simulations, spherical-like coordinates are more suitable for nearly spherical remnants and azimuthal flows due to lower numerical dissipation in the evolution of fluid angular momentum, as well as requiring fewer numbers of computational cells. However, the use of spherical coordinates to numerically solve hyperbolic partial differential equations can result in severe Courant-Friedrichs-Lewy (CFL) stability condition time step limitations, which can make simulations prohibitively expensive. This paper addresses this issue for the numerical solution of coupled spacetime and general relativistic magnetohydrodynamics evolutions by introducing a double fast Fourier Transform (FFT) filter and implementing it within the fully message passing interface ({\sc MPI})-parallelized \SphericalNR framework in the {\sc Einstein Toolkit}. We demonstrate the effectiveness and robustness of the filtering algorithm by applying it to a number of challenging code tests, and show that it passes these tests effectively, demonstrating convergence while also increasing the
 time step significantly compared to unfiltered simulations.
\end{abstract}

 \maketitle

\section{Introduction}\label{sec:introduction}

With the advent of gravitational wave and multimessenger astronomy~\cite{Abbott:2016blz,
TheLIGOScientific:2016wfe,Abbott:2016nmj,Abbott:2017vtc,Abbott:2017gyy,Abbott:2017oio,TheLIGOScientific:2017qsa,
GBM:2017lvd,Monitor:2017mdv}, there is an ever greater need for high-accuracy, long-term numerical simulations of merging compact objects and their remnants, such as the first general 
relativistic hydrodynamics (GRHD) binary neutron star (NS) merger 
simulation~\cite{ShibataBNS2000}, the first simulations of binary black hole (BH) 
mergers~\cite{Pretorius2005,Campanelli2006,Baker2006},
the first GRHD BH-NS merger simulation~\cite{Shibata2006},
the first general relativistic magnetohydrodynamics (GRMHD) BNS merger
simulations~\cite{HAD2008,Liu2008}, and the first GRMHD simulation of BH-NS
mergers~\cite{ChawlaBHNS2010}. See the review articles~\cite{2012LRR....15....8F,2014ARA&A..52..661L,2017RPPh...80i6901B,2017ARNPS..67..253T,2019LRR....23....1M,2019RPPh...82a6902D,2019ARNPS..69...41S,2019PrPNP.10903714B,2021LRR....24....5K} and references therein for recent advances in the field. Traditionally, such simulations 
are performed using Cartesian coordinates, which leads to simpler numerical algorithms and very 
robust codes. However, such coordinates are also computationally wasteful, as they over-resolve 
in the angular directions leading to the necessity of mesh refinement in order to prevent 
computationally prohibitive cell counts in large computational domains. 

An alternative approach is to use coordinates adapted to the symmetries (approximate or exact) associated with the numerical problem. In particular, the nearly spherical remnant associated with a compact-object merger is ideally suited for spherical-like coordinates due to the lower numerical dissipation in the evolution of fluid angular momentum compared to Cartesian coordinates~\cite{2022PhRvD.106h3015L}. Another area are GRMHD simulations of accretion disks, where it is 
customary to use spherical-like coordinates (see, for instance, the Einstein Horizon Telescope code comparison project~\cite{2019ApJS..243...26P}). With this in mind, we recently introduced SphericalNR~\cite{Mewes:2018szi}, a fully {\sc MPI}-parallelized implementation of the Baumgarte-Shapiro-Shibata-Nakamura (BSSN)~\cite{Shibata:1995we, Baumgarte:1998te} formulation of the Einstein equations in spherical coordinates within the \ET\footnote{\href{https://einsteintoolkit.org/}{https://einsteintoolkit.org/}}~\cite{2012CQGra..29k5001L}. The code was later extended to include GRMHD~\cite{Mewes:2020vic} in the reference metric formalism~\cite{2014PhRvD..89h4043M} and constraint damping in the spacetime evolution
via the fully covariant and conformal formulation of the Z4 system, fCCZ4~\cite{AlicCCZ4,AlicCCZ4_matter,fCCZ42014,Mewes:2020vic}.

The attractive features of using spherical coordinates for the simulation
of azimuthal flows comes with a price, however, as the use of spherical coordinates can lead to a severe Courant-Friedrichs-Lewy stability condition (CFL)~\cite{CFL_1928} limitation of the allowable time step associated with the polar axis and origin of the spherical coordinate system when solving hyperbolic partial differential equations. This is due to the cell volumes (and therefore time steps) becoming prohibitively small as the polar axis and origin are approached. Compared to Cartesian coordinates, where
the time step is $\propto dx_{\rm min}$, in spherical coordinates the time step is 
$\propto r \sin\theta d\varphi$, which can render high resolution, long-term numerical simulations in 3D prohibitively expensive. There are various approaches to remedy the problem, including multiblock or multipatch techniques~\cite{Ronchi1996,Gomez:1996ge,1997PhRvD..56.6298B,Gomez:1998uj,1999bhgr.conf..383B,Kageyama2004,Thornburg:2004dv, Diener:2005tn, Lehner:2005bz, Schnetter:2006pg, Zink:2007xn, 2008SIAMR..50..723C,Pollney:2009yz,Fragile2009a,Wongwathanarat2010,Melson2015,ShiokawaPW2018,2020ApJS..248...11B}, tessellated grids~\cite{1968MWRv...96..351S,2003CMAME.192.3933D}, static mesh refinement~\cite{Mueller2015,2018MNRAS.474L..81L,2019ApJS..241....7S}, mesh coarsening~\cite{2019MNRAS.484.3307M,2019PASJ...71...98N,2019JCoPh.376..276Z,2021GMD....14..859D}, local filters~\cite{Shapiro1970,Gent1989,Jablonowski2004}, global FFT filters~\cite{2019MNRAS.484.3307M}, and distorted angular grids~\cite{Korobkin2011,Noble2012}, to name a few. Each approach to solve 
the CFL limitation has its own advantages and limitations, such as algorithmic 
complexity, ensuring conservation, or the use of global operations.

In {\sc SphericalNR}, we have chosen to implement a double FFT filter that filters spacetime and GRMHD fields in both the $\theta$ and $\varphi$ directions depending on radius and latitude (for the filtering in $\varphi$). FFT filtering has both conceptual and algorithmic difficulties: In general, the evolved GRMHD fields can develop discontinuities, which requires a different filter algorithm than filtering smooth fields by exponentially damping CFL unstable modes. Further, the FFT filter is a global operation of either an entire great circle when filtering in $\theta$ or an entire $\varphi$ coordinate ring. An earlier version of the FFT filter that was only {\sc OpenMP}-parallelized and filtering in the $\varphi$ coordinate only was used in~\cite{Mewes:2020vic,2023ApJ...947L..34M}. This severely limited the applicability of the filter to high-resolution simulations due to the inability to decompose the domain in $\varphi$. In this work, we
have extended the FFT filter to work in both angular coordinates and have fully {\sc MPI}-parallelized it. We have developed an automatic switch to filter the GRMHD fields
with a Gaussian filter instead of an exponential filter which prevents spurious oscillations
as a result of filtering discontinuous fields.

This paper is organized as follows. In Sec.~\ref{sec:techniques}, we describe the techniques we use to both evolve the BSSN/fCCZ4 system coupled to GRMHD and how we filter the unstable polar and azimuthal modes in the double FFT filter, as well as describing the details of the filter parallelization. In Sec.~\ref{sec:results}, we show the results of applying our filtering algorithm to a single spinning Bowen-York black hole (BH), an off-center spherical explosion, an off-center stable rotating neutron star (NS), and a rotating NS that is susceptible to the
dynamical bar-mode instability. Finally, in Sec.~\ref{sec:discussion}, we discuss our results. We use the Einstein summation convention throughout. Unless otherwise stated, all results are presented in units in which $G=M_{\odot}=c=1$.

\section{Techniques}\label{sec:techniques}

In previous papers, our collaboration described a fully parallelized implementation of the 
vacuum Einstein equations and GRMHD using spherical coordinates~\cite{Mewes:2018szi,Mewes:2020vic} within the 
{\sc Einstein Toolkit}. Here, we describe a series of modifications that allow us to use that code without the sometimes severe CFL limitation on the time step. Our code is based on the fCCZ4 formalism of Einstein equations and the {\it Valencia} formulation of GRMHD~\cite{1997ApJ...476..221B,2006ApJ...637..296A} and uses the \ET to provide parallelization and critical analysis tools. The \ET is an open-source code suite for relativistic astrophysics simulations. It uses the modular {\sc Cactus}\footnote{\href{https://www.cactuscode.org}{https://www.cactuscode.org}} framework~\cite{Goodale02a} (consisting of general modules called \enquote{thorns}) and provides adaptive box-in-box mesh refinement (AMR) via the {\sc Carpet}\footnote{\href{https://bitbucket.org/eschnett/carpet}{https://bitbucket.org/eschnett/carpet}} code~\cite{2004CQGra..21.1465S}.

In the present work, we introduce two main modifications to the standard evolution techniques described in~\cite{Mewes:2018szi,Mewes:2020vic}, these are the introduction of a double FFT filtering scheme to ameliorate the severe CFL limitations associated with spherical coordinates and a generic fisheye~\cite{Baker:2001sf}  radial coordinate to more efficiently allocate the grid points (we also introduce modifications to the standard shift conditions that appears to perform better in some of our tests).
While we do choose a few particular ``fisheye'' coordinates here, for example
\begin{equation}
  r = A x_1 + (1-A) r_0 {\rm atan}\left(x_1/r_0\right),
\end{equation}
where $r$ is the usual radial coordinate, $x_1$ is the \enquote{fisheye} radial coordinate (the actual numerical coordinate), the constant $A$ determines the ratio of the physical to numerical gird spacing far from the origin (by construction, this ratio is 1 at the origin), and $r_0$ is a parameter to fine-tune where the transition occurs, the code can work with any one-to-one differentiable function $r(x_1)$. In particular, we performed several simulations with an {\it exponential} ``fisheye'', $r(x_1) \propto \exp(x_1)$, commonly used in GRMHD simulations of accretion disks (see, e.g. some of the
codes used in~\cite{2019ApJS..243...26P}).

We give a summary of the evolution system below, and refer the reader to the full details in~\cite{Baumgarte:2012xy, Baumgarte:2015dya, Ruchlin:2017com, Mewes:2020vic}. Central to the method is  the conformally related spatial metric
\begin{equation}
\bar{\gamma}_{ij} = e^{-4\phi} \gamma_{ij},
\end{equation}
where $\gamma_{ij}$ is the physical spatial metric, and $\phi$ the conformal factor 
\begin{equation}
e^{4\phi} = (\gamma / \bar{\gamma})^{1/3},
\end{equation}
where $\gamma$ and $\bar{\gamma}$ are the determinants of the physical and conformally related metric, respectively. In order to make the conformal rescaling unique, we adopt Brown's \enquote{Lagrangian} choice~\cite{Brown:2009dd}
\begin{equation}
\partial_t \bar{\gamma} = 0,
\end{equation}
fixing $\bar{\gamma}$ to its initial value throughout the evolution. Similarly, the conformally related extrinsic curvature is defined as
\begin{equation}
\bar{A}_{ij} = e^{-4\phi}\left(K_{ij}-\frac{1}{3}\gamma_{ij}K\right),
\end{equation}
where $K_{ij}$ is the physical extrinsic curvature and $K=\gamma^{ij}K_{ij}$ its trace.

The main idea is to write the conformally related metric as the sum of the flat background metric plus perturbations (which need not be small)
\begin{equation}
\bar{\gamma}_{ij} = \hat{\gamma}_{ij} + \epsilon_{ij},
\end{equation}
where $\hat{\gamma}_{ij}$ is  the reference metric in fisheye spherical coordinates,
\begin{equation}
\hat{\gamma}_{ij} =
\begin{pmatrix}
\left(\frac{dr(x_1)}{dx_1} \right)^2 & 0 & 0 \\
0 & r(x_1)^2 & 0 \\
0 & 0 & r(x_1)^2 {\rm sin}^2 \theta
\end{pmatrix}.
\end{equation}
The conformal connection coefficients $\bar{\Lambda}^i$ are treated as independently evolved variables that satisfy the initial constraint
\begin{equation}
\bar{\Lambda}^i - \Delta^i = 0.
\end{equation} 
Here 
\begin{equation}
\Delta^i \equiv \bar{\gamma}^{jk}\Delta^i_{jk}
\end{equation}
and $\Delta^i_{jk}$ is the difference between the Christoffel symbols of the conformally rescaled and flat reference metric,
\begin{equation}
\Delta^i_{jk} \equiv \bar{\Gamma}^i_{jk} - \hat{\Gamma}^i_{jk}.
\end{equation}
The conformal connection coefficients $\bar{\Lambda}^i$, therefore, transform like vectors in the reference-metric formalism.
Together with the lapse $\alpha$ and the shift $\beta^i$, this set of the $3+1$ variables $\{\alpha,\beta^i,\gamma_{ij},K_{ij}\}$, expressed in spherical coordinates, is stored in the thorn {\sc ADMBase} to interface with existing diagnostics in the \ET. 

When evolving the fCCZ4 (and other systems similar to BSSN) equations, we use the $1+\log$ slicing conditions~\cite{1+log_Bona1995}
\begin{equation}
  \partial_t \alpha = \beta^i \partial_i \alpha - 2 \alpha K,
\end{equation}
and shift conditions~\cite{Gamma-driver2003,Brown:2009dd} 
\begin{align}
  \partial_t B^i &= \frac{3}{4} \partial_t \bar\Lambda^i - \kappa_B \frac{3}{4} \beta^j \hat D_j \bar\Lambda^i - \eta B^i, \nonumber\\
  \partial_t \beta^i &= B^i, \label{eq:someshift}
\end{align}
where $\hat D_i$ is the covariant derivative with respect to the background flat metric and $\kappa_B=0$ leads to the standard nonadvected $\Gamma$-driver shift, while $\kappa_B=1$ leads to a modification that proved to be more accurate for spacetimes containing BH.

A key idea for regularizing the fCCZ4 (and other) systems in spherical coordinates is to evolve tensorial quantities in a basis that is orthonormal with respect to the background conformal metric. To distinguish between coordinate-basis components and orthonormal-basis components, we will follow the notation of~\cite{Mewes:2020vic}. Suppose ${T^{i_1 i_2 \cdots}}_{j_1 j_2 \cdots}$ are the coordinate components of a tensor $T$, then the orthonormal components will be denoted by $ {T^{\{a_1\}\{a_2\}\cdots}}_{\{b_1\}\{b_2\}\cdots}$, where
\begin{eqnarray}
   && {T^{\{a_1\}\{a_2\}\cdots}}_{\{b_1\}\{b_2\}\cdots}  \nonumber\\
   &= &{\bf e}^{\{a_1\}}_{i_1} {\bf e}^{\{a_2\}}_{i_2}\cdots {\bf e}^{j_1}_{\{b_1\}} {\bf e}^{j_2}_{\{b_2\}}\cdots {T^{i_1 i_2\cdots}}_{j_1 j_2\cdots},
\end{eqnarray} 
and ${{\bf e}}^{\{j\}}_{i}$ and ${{\bf e}}^{i}_{\{j\}}$ are elements of the (background) orthonormal vector and covector bases, respectively. In our notation ${\bf e}^{\{i\}}_{j}$ represents the $j$th coordinate component of the $i$th basis element.

The background orthonormal vector basis takes the form
\begin{eqnarray}
  {\bf e}^j_{\{r\}} &=& \left( \frac{1}{r'(x_1)},0, 0\right),\\
  {\bf e}^j_{\{\theta\}} &=& \left( 0, \frac{1}{r(x_1)}, 0\right),\\
  {\bf e}^j_{\{\varphi\}} &=& \left( 0, 0,\frac{1}{r(x_1)\ \sin\theta}\right),
\end{eqnarray}
with the corresponding orthonormal cobasis,
\begin{eqnarray}
  {\bf e}_j^{\{r\}} &=& \left(r'(x_1),0, 0\right),\\
  {\bf e}_j^{\{\theta\}} &=& \left( 0, r(x_1), 0\right),\\
  {\bf e}_j^{\{\varphi\}} &=& \left( 0, 0, r(x_1)\ \sin\theta \right).
\end{eqnarray}

In this system, our evolution variables are $\epsilon_{\{i\}\{j\}}$, $A_{\{i\}\{j\}}$, etc.. To convert the coordinate-component evolution equation to the orthonormal-basis components, we express derivative of the coordinate-component tensors in terms of analytical derivatives of the basis and finite-difference derivatives of the tensor components. For example, an expression like 
\begin{equation}
  \partial_i A_{jk}
\end{equation}
becomes
\begin{eqnarray}
  && \partial_i \left( {\bf e}_j^{\{l\}} {\bf e}_k^{\{m\}} A_{\{l\}\{m\}} \right) \nonumber \\
  &=& \partial_i \left( {\bf e}_j^{\{l\}} {\bf e}_k^{\{m\}} \right)  A_{\{l\}\{m\}} + {\bf e}_j^{\{l\}} {\bf e}_k^{\{m\}} \partial_i   A_{\{l\}\{m\}},
\end{eqnarray}
where $ \partial_i   A_{\{l\}\{m\}}$ is evaluated using finite differences and the derivatives of the basis elements are calculated analytically.

The numerical code for the right-hand side in the fCCZ4 evolution system as well as GRMHD source terms
are provided by the \SENR code, and the time integration is performed with the method of lines as implemented in the {\sc MoL}~\cite{2012CQGra..29k5001L} thorn. We have implemented a fourth-order strong stability-preserving Runge-Kutta (SSPRK54) method~\cite{spiteri2002new} in {\sc MoL}, which has larger CFL factor than the more traditional RK2 or RK3~\cite{ShuOsher1988, GottliebTVDRK1998} methods and is large enough to compensate for the extra computational work due to SSRK54 having more stages.

We refer the reader to~\cite{Baumgarte:2012xy, Baumgarte:2015dya, 
Ruchlin:2017com,Mewes:2018szi,Mewes:2020vic} for the full details of the evolution system but 
note that, compared to~\cite{Mewes:2020vic}, we have made several improvements to the GRMHD code
in {\sc SphericalNR}: We have developed a custom built ninth order WENO-Z9 reconstruction 
(local smoothness indicators $\beta_k$ written as perfect squares~\cite{2016JCoPh.326..780B}, 
optimal higher order global smoothness indicators $\tau_{r2-1}$~\cite{2011JCoPh.230.1766C}, and 
adaptive 
$\epsilon$~\cite{2007MNRAS.379..469T}). There is also the option to combine the WENO-Z9 
reconstruction with the monotonicity-preserving (MP) limiter~\cite{MP5_1997}, 
resulting in a MPWENO scheme~\cite{2000JCoPh.160..405B}. We have implemented the consistency-ensuring summation of~\cite{fleischmann2019numerical}, which when
applied to the MP limiting algorithm helps
alleviating the spontaneous symmetry breaking and associated drifts we observed 
in~\cite{Mewes:2020vic} when using MP5 reconstruction. We have also implemented seventh and 
ninth order MP7 and MP9~\cite{MP5_1997}, but find that, even using 
consistency-ensuring summation, MPWENO-Z9 is still more robust in that regard (and WENO-Z9 better still). 
When using higher order reconstruction methods, the reconstructed density or pressure might 
occasionally become negative, in which case we reconstruct them using a total variation 
diminishing (TVD) reconstruction with the minmod limiter. 
We have also implemented higher order flux corrections of~\cite{ECHO_code2007}
using cell-centered fluxes as higher order corrections to face fluxes as 
in~\cite{2016JCoPh.305..604C}. The WENO-Z9 method will be described in detail in 
a forthcoming paper regarding the use of higher order methods in GRMHD 
simulations of BH accretion flows.

When using WENO-Z9 in simulations (we can still use any of the existing reconstruction 
methods available in the original {\sc GRHydro} code~\cite{GRHydro_code_paper}), we typically 
evolve the magnetic vector potential $A_i$ and the electromagnetic scalar potential 
$\hat{\Phi}$ using 
tenth order central finite differences and use ninth order Kreiss-Oliger (KO) 
dissipation~\cite{kreiss1973methods} to damp high frequency noise in the evolution of $A_i$ and 
$\hat{\Phi}$. We obtain the magnetic field $B^i$ from $A_i$ using tenth order finite differences 
when calculating the curl of $A_i$. Accordingly, we use tenth ordered central finite 
differences in the source terms [Eqs. (73) and (80) in~\cite{Mewes:2020vic}].
Unless otherwise noted in Sec.~\ref{sec:results} below, we have used WENO-Z9 
and tenth order central finite differences.

We note that these higher-order methods do require more ghost zones, which can have an impact on speed. A tenth-order central stencil requires five ghost zones.

Finally, in order to improve the robustness of the GRMHD evolution, we have also made several 
improvements, in particular to the primitive variable recovery and artificial atmosphere.
Before attempting primitive recovery, we enforce the following condition on the conserved variables found in Appendix C of~\cite{2012PhRvD..85f4029E}:
\begin{equation}
    \tau = \max\left(\tau, \sqrt{\gamma} \left((\rho \epsilon)_{\mathrm{floor}} + \frac{B^2}{2} \right) \right),
\end{equation}
as well as steps (2) and (3) of said Appendix.
We then use the primitive variable recovery scheme of~\cite{2006ApJ...641..626N}. If the initial
recovery fails, we try again using the initial guesses of~\cite{CerdaDuran:2008pv}. In the regions where the primitive variable recovery becomes increasingly difficult (low plasma-$\beta$ $2 P/b^2$, high internal energy density $\epsilon$, high Lorentz factor $W$), the primitive recovery is still prone to fail. To alleviate this problem, we follow~\cite{1999JCoPh.148..133B,2009ApJ...692..411N} and evolve the conserved entropy 
in the reference metric formalism,
\begin{equation}\label{eq:ee_eq}
    \partial_t \mathcal{S} + \hat{\mathcal{D}}_i \left(\alpha \mathcal{S} \left(v^i - \frac{\beta^i}{\alpha}\right)\right)= 0,
\end{equation}
where 
\begin{equation}
    \mathcal{S} = e^{6 \phi} \sqrt{\frac{\bar{\gamma}}{\hat{\gamma}}}W\frac{P}{\rho^{\Gamma -1}},
\end{equation}
and $\hat{\mathcal{D}}$ is the covariant derivative associated with the spherical background 
metric $\hat{\gamma}_{ij}$, $\alpha$ the lapse, $v^i$ the Valencia fluid three-velocity, 
$\beta^i$ the shift, $\phi$ the conformal factor, $\bar{\gamma}$ the determinant of the 
conformal metric $\bar{\gamma}_{ij}$, $\hat{\gamma}$ the determinant of the spherical 
background metric, $P$ the fluid pressure, $\rho$ the fluid rest-mass density, and $\Gamma$ the 
adiabatic index, respectively (see~\cite{Mewes:2020vic} for details on the evolution
equations in the reference metric formalism). We always use a 
TVD reconstruction with the minmod limiter for the reconstruction of the entropy. After each 
successful primitive recovery, $\mathcal{S}$ is recalculated from the primitives and evolved 
for a Runge-Kutta substep. We recover the pressure $P$ from $\mathcal{S}$ wherever 
$\beta^{-1} = b^2/(2 P) > 100$ or using the recovery scheme of~\cite{2006ApJ...641..626N} failed. While this approach 
guarantees a positive pressure, the recovery can still fail, in which case we 
follow~\cite{2009ApJ...692..411N} and try to average the primitives from neighboring cells that 
had a successful recovery; otherwise, the primitives are set to atmosphere values with $v^i=0$. The magnetic field $B^i$ is never touched and always calculated from the curl of $A_i$.

For the artificial atmosphere, we have implemented both isotropic and radially dependent floors for the density and pressure, where 
\begin{align}
    \rho_{\rm floor} &= \rho_{\rm atmo}\, \max(r_{\mathrm{min}},r)^{-1.5} \\
    P_{\rm floor} &= (\Gamma - 1) (\rho \epsilon)_{\rm  floor} \\ \nonumber
    &= (\Gamma - 1) (\rho \epsilon)_{\rm atmo}\, \max(r_{\mathrm{min}},r)^{-1.5\, \Gamma},
\end{align}
where $r_{\mathrm{min}}$ is a parameter to avoid the floors from diverging at the origin. Where evolved cells fall below these floor values we just raise $\rho$ or $P$ to their floor values, and if $10 b^2 > {\rm min}(\rho,p/(\Gamma-1))$, we add matter in the drift frame~\cite{2017MNRAS.467.3604R} instead. When using the approximate HLLE (Harten-Lax-van Leer-Einfeldt) Riemann solver~\cite{EinfeldtHLLE1988,HartenHLLE1983}, we switch to the more diffusive global (with a characteristic speed of 1) Lax-Friedrichs fluxes wherever the magnetization $\sigma = \frac{b^2}{\rho} > 1 $, the inverse plasma $\beta^{-1} > 100$,  $\rho < 10 \, \rho_{\rm floor}$, $\rho \epsilon < 10 \, (\rho \epsilon)_{\rm floor}$, or a grid point is inside an apparent horizon. The entropy equation~\eqref{eq:ee_eq} is always evolved with the Lax-Friedrichs flux and a global characteristic speed of 1. We also impose a ceiling (typically 50) on W, and a ceiling on 
$\epsilon$. After these potential fixes to the primitive variables are done, we 
recompute the conserved variables everywhere. While this breaks strict conservation,
it is necessary to maintain a consistent set of conserved and primitive variables. 

Finally, inspired by other GRMHD codes~\cite{BHAC2017,2018MNRAS.474L..81L}, at the 
$r=0$ cell faces we set the reconstructed electric field components 
$E_{\theta}=E_{\varphi}=0$, and for the $\theta = 0,\pi$ cell faces set $E_{r} = E_{\varphi} = 0$.

\subsection{Filtering algorithms}

We use the {\sc FFTW3}~\cite{FFTW3} library to perform all Fourier transforms. 
The main idea of the algorithm is to dampen CFL unstable modes at a given radius and latitude by performing FFTs in both polar and azimuthal directions, modifying the Fourier expansion of the evolved fields, and then performing the inverse FFT to obtain the filtered fields in real space. The double FFT filter first performs FFT filtering in the $\theta$ direction followed by FFT filtering in the $\varphi$ direction. In order to be able to filter in the $\theta$ direction, we define a new angular coordinate, $\vartheta$, which extends the $\theta$ coordinate from $[0,\pi]$ to $[0,2 \pi]$. To do this, we first construct the field ${\bf X}(x_1, \vartheta, \varphi)$,
\begin{equation}
    {\bf X}(x_1, \vartheta, \varphi)=
    \begin{cases}
      {\bf X}(x_1, \theta, \varphi), & \vartheta \in [0,\pi]\\
      (-1)^a{\bf X}(x_1, \pi-\theta, \pi + \varphi), & \vartheta \in [\pi,2 \pi]
    \end{cases}
\end{equation}
where $a=0$  or 1, depending on the axis parity factor of the field, i.e., positive or negative parity, respectively (see Table I in~\cite{Mewes:2020vic}).
We then perform a FFT in the $\vartheta$ coordinate on ${\bf X}(x_1, \vartheta, \varphi)$ to obtain the Fourier expansion ${\bf \tilde{X}}(x_1,l,\varphi)$ (note that $l$ denotes Fourier mode in the $\theta$ direction), which is then filtered and finally obtain the filtered field ${\bf X}(x_1, \vartheta, \varphi)$ by performing the inverse FFT,
\begin{align}
    {\bf X}(x_1, \vartheta, \varphi) &\xrightarrow[]{\rm FFT} {\bf \tilde{X}}(x_1,l,\varphi) 
    \to f(l,l_{\mathrm{max}}) {\bf \tilde{X}}(x_1,l,\varphi) \nonumber \\
    &\xrightarrow[]{\rm iFFT} {\bf X}(x_1, \vartheta, \varphi),
\end{align}
where the filtering function $f(l,l_{\mathrm{max}})$ depends on the type of field being filtered and will be described below (see~\eqref{exp_filter} and \eqref{gaussian_filter}). We then filter the evolved fields in the $\varphi$ direction analogously,
\begin{align}
    {\bf X}(x_1, \theta, \varphi) &\xrightarrow[]{\rm FFT} {\bf \tilde{X}}(x_1,\theta,m) 
    \to f(m,m_{\mathrm{max}}) {\bf \tilde{X}}(x_1,\theta,m) \nonumber \\
    &\xrightarrow[]{\rm iFFT} {\bf X}(x_1, \theta, \varphi)
\end{align}
(note that $m$ denotes Fourier modes in the $\varphi$ direction).
The maximum allowed modes $l_{\rm max}$ and $m_{\rm max}$ in the $\theta$ and $\varphi$ filters are given by,
\begin{align}
    l_{\rm max} &= \max \left(2, \frac{2\, r}{dr_{\rm min}} {\cal L} \right),\\
    m_{\rm max} &= \max \left(2, \frac{2 r}{dr_{\rm min}} \sin{\theta} {\cal L} \right),
\end{align}
where $r$ is the physical coordinate radius (i.e., related to the compuational radial coordinate by a fisheye transformation) and $dr_{\mathrm{min}}$ is the smallest radial grid spacing on the computational domain. Because even along the pole and the origin the angular dependence of the evolved fields are nontrivial, we never filter out the first ${\cal L}$  modes (see discussion below).  

If we would like to achieve a time step that is $\propto dr_{\rm min}$, we would need to filter the
evolved fields to $m_{\rm max}$ near the axis for even moderate angular resolutions.
However, we can never (in, general) filter all the way to $m=0$ and $l=0$. In the vicinity of both the poles and the origin, a regular metric in Cartesian coordinates will induce both $m=1$ and $m=2$ modes in the resulting spherical metric (in particular, the components in the orthonormal basis). This is easily shown by considering a generic metric in Cartesian coordinates in the vicinity of the pole (i.e., $x\sim0$, $y\sim 0$, and $z=z_0$). The metric will, in general, be
\begin{eqnarray}
  ds^2 &=& (1+a) dx^2 + 2 b\ dx dy + 2 c dx dz + (1+d) dy^2 \nonumber \\ 
  &+& 2 e\ dy dz + (1+f) dz^2 \nonumber \\
  &+& {\cal O}(x,y, z - z_0).
\end{eqnarray}
The resulting components of the metric in the background (spherical) orthonormal basis will contain terms proportional to $\exp(\pm i \varphi)$, $\exp(\pm i 2 \varphi)$, $\exp(\pm i \theta)$ and $\exp(\pm i 2 \theta)$. For example,
\begin{eqnarray}
  \gamma_{\{r\}\{r\}} &=& (a+1) \sin ^2(\theta ) \cos ^2(\varphi )\nonumber \\
  &&+\sin (\varphi ) (\sin ^2(\theta ) (2 b \cos (\varphi )+(d+1)
   \sin (\varphi ))\nonumber \\
   &&+e \sin (2 \theta ))+c \sin (2 \theta ) \cos (\varphi )\nonumber \\
   &&+(f+1) \cos ^2(\theta ),\\
  \gamma_{\{\varphi\}\{\varphi\}} &=& 1/2 (2 + a + d + (-a + d) \cos[2 \varphi]\nonumber\\
  &&- 2 b \sin[2 \varphi]),\\
  \gamma_{\{\theta\}\{\varphi\}} &=&\cos (\theta ) ((d-a) \sin (\varphi ) \cos (\varphi )\nonumber \\
  &&+b \cos (2 \varphi ))+\sin (\theta ) (c \sin (\varphi )-e \cos (\varphi ))
\end{eqnarray}
On the poles, these become
\begin{eqnarray}
\gamma_{\{r\}\{r\}} &=& 1+f, \\
\gamma_{\{\varphi\}\{\varphi\}} &=&1 + ((d-a) \cos (2 \varphi )+a \nonumber \\
&-&2 b \sin (2 \varphi )+d)/2,\\
\gamma_{\{\theta\}\{\varphi\}} &=& (d-a) \sin (\varphi ) \cos (\varphi )+b \cos (2 \varphi ).
\end{eqnarray}
Thus, at the origin, and on the poles, there should be $m=2$ modes in both $\theta$ and $\varphi$ (but no higher) if the metric in spherical coordinates is to reproduce this simple Cartesian metric.

To fully retain the $m=2$ modes in both $\theta$ and $\varphi$, we could set all unwanted modes to zero (a spectrally sharp low-pass filter essentially) or use an exponential filter:
\begin{equation}\label{exp_filter}
    f_{\mathrm{exp}}(l,l_{\mathrm{max}}) = \left\{\begin{array}{lr}  1, & |l| \leq l_{\rm max} \\
  e^{-(|l| - l_{\rm max})}  , & l > l_{\rm max} \end{array} \right.
\end{equation}
which retains all power in modes $|l| \leq l_{\rm max}$. We use the exponential filter for all evolved spacetime fields which are smooth
\footnote{The $1+\log$~\cite{1+log_Bona1995} and \enquote{$\Gamma$-driver}~\cite{Gamma-driver2003} gauge conditions for the evolution of the lapse and shift we use in our evolution can actually develop true shocks (see~\cite{1997PhRvD..55.5981A,1998PhRvD..57.4511A,2005CQGra..22.4071A,2003CQGra..20..607A} for a description of the pathologies and gauge conditions that are shock-avoiding), but we haven't seen any evidence for the appearance of such gauge shocks and related problems with using the exponential filter for the spacetime fields in our simulations. The shock-avoiding slicing conditions of~\cite{1997PhRvD..55.5981A,2003CQGra..20..607A} were shown to be a viable alternative to the $1+\log$ slicing condition~\cite{2022PhRvD.106d4014B}, which means they could be used in conjunction with the exponential filter in situations that are prone to the development of gauge shocks.}
and retaining full power in unfiltered modes therefore does not result in new 
extrema and potential Gibbs phenomenon during the filtering. This situation is very 
different for the filtered GRMHD fields $(D, S_i, \tau, \mathcal{S}, A_i, 
\hat{\Phi})$, which can in general become discontinuous. Filtering discontinuous fields with the 
exponential filter would result in Gibbs phenomenon, leading to  nonphysical 
oscillations and potentially nonpositive values (the latter will result in 
catastrophic failures that would need to be fixed in an posterior step after 
filtering, similar to what is done in the artificial atmosphere). To remedy this, 
we filter $(D, S_i, \tau, \mathcal{S})$ with a Gaussian filter:
\begin{equation}\label{gaussian_filter}
    f_{\mathrm{Gaussian}}(l,l_{\mathrm{max}}) = \exp \left(\log(0.9)\left(\frac{l}{l_{\rm max}+1} \right)^2 \right),
\end{equation}
while filtering the magnetic vector potential and electromagnetic scalar potential $(A_i, \hat{\Phi})$ with the exponential filter~\eqref{exp_filter}. The Gaussian filter avoids Gibbs 
phenomena at the expense of reducing the power in all Fourier modes apart from the $l,m=0$ 
mode. While this guarantees that no new extrema are generated in the filtering process, the 
reduction in power in lower order modes negatively affects the overall resolution of the simulation.
We have 
chosen a reduction in power to 0.9 in the first mode that is CFL unstable (i.e., the mode $l = l_{\rm max} + 1$) as a compromise between 
trying to reduce as little power as possible in the modes that are CFL stable (and should 
therefore not be filtered at all) while simultaneously guaranteeing the CFL unstable modes are 
 sufficiently suppressed to prevent CFL instabilities in the evolution.

In our initial explorations of the tests presented in Sec.~\ref{sec:results}, we found that maintaining full power in the modes up to $m=2$ at the axis and origin are critical. We have therefore designed a hybrid filter that seeks to use the exponential filter wherever possible and switches to the Gaussian filter only when discontinuities are detected in the coordinate ring being filtered. For a field $u$, we try to detect discontinuities in the $\vartheta$ and $\varphi$ rings using Jameson's shock detector~\cite{jameson1981numerical}:
\begin{equation}
    \sigma_i = \frac{\left| u_{i-1} -2 u_i + u_{i+1}\right|}{|u_{i-1}|+2|u_i|+|u_{i+1}|+\epsilon}, i={\theta,\varphi},
\end{equation}
where $u_i, u_{i\pm1}$ represent the field value in the current cell and its neighbours, and $\epsilon$ is a small number used to avoid the division by zero in the denominator.
If a single cell in a $\vartheta$ or $\varphi$ ring fulfils $\sigma_i >= 0.95$, we use the Gaussian filter for a given ring and field; otherwise, we use the exponential filter.

Finally, we note that the characteristic speeds for the GRMHD evolution are always less than the characteristic speeds of the spacetime evolution (some gauge modes have speeds of $\sqrt{2}$), and the GRMHD evolution therefore typically allows for larger CFL factors than the evolution of the fCCZ4 variables. We therefore typically choose ${\cal L}=3$ for the metric fields, and ${\cal L}=12$ for the matter fields. 
 
\subsubsection{Filtering spacetime fields in the presence of BHs}
The spacetime evolution in {\sc SphericalNR} is subject to two algebraic constraints:
\begin{align}
   \bar{\gamma} &=\hat{\gamma}\label{detg_constraint} \\ 
   \bar{\gamma}^{ij}\bar{A}_{ij} &= 0, \label{trA_constraint}
\end{align}
where $\bar{\gamma}$ and $\hat{\gamma}$ are the determinants of the conformally
related metric $\bar{\gamma}_{ij}$ and the background metric $\hat{\gamma}_{ij}$, and 
$\bar{A}_{ij}$ is the conformally related extrinsic curvature, respectively.
In the development of the double FFT filter, we noticed that we need to adjust
the way these two constraints are enforced when filtering in the $\theta$ direction in the 
presence of BH spacetimes in order to obtain a stable evolution.
Usually, at each Runge-Kutta substep in the evolution, we enforce the above constraints
by making the following substitutions at all grid points in the domain:
\begin{equation}
    h_{\{i\}\{j\}} \to \left( \frac{\hat{\gamma}}{\bar{\gamma}} \right)^{\frac{1}{3}} \left(\delta_{ij} + h_{\{i\}\{j\}} \right) - \delta_{ij},
\end{equation}
where $\delta_{ij}$ is the Kronecker delta, and
\begin{equation}
    \bar{A}_{\{i\}\{j\}} \to \bar{A}_{\{i\}\{j\}} - \left(\delta_{ij} + h_{\{i\}\{j\}} \right) \frac{\bar{\gamma}^{ij}\bar{A}_{ij}}{3},
\end{equation}
where we again follow the notation of~\cite{Mewes:2020vic}, namely that indices in curly 
braces represent components in the orthonormal basis with respect to the spherical background 
metric, whereas normal indices represent components in the coordinate basis.

Enforcing the algebraic constraints this way turned out to be unstable 
in the presence of BH spacetimes when filtering in the $\theta$ direction
close to the center of a BH (regardless
if we evolve initial data containing BH initial data or in situations where 
a BH is dynamically formed during evolution, such as the collapse of a NS).

To remedy this instability, we adapt the way we enforce the algebraic constraints as follows: in regions where we need to apply the filtering in the 
$\theta$ direction for CFL stability and the lapse $\alpha < 0.3$, we enforce the
constraint~\eqref{detg_constraint} as follows (see also~\cite{2002PhRvD..66h4026Y}):
\begin{align}
    h_{\{1\}\{1\}} &= \left(\hat{\gamma} + \bar{\gamma}^2_{12} \bar{\gamma}^{}_{33}
    -2 \bar{\gamma}^{}_{12} \bar{\gamma}^{}_{13} \bar{\gamma}^{}_{23} 
    + \bar{\gamma}^2_{13} \bar{\gamma}^{}_{22} \right) \nonumber \\
    &/\left(\bar{\gamma}^{}_{22} \bar{\gamma}^{}_{33} - \bar{\gamma}^2_{23}\right) / (\hat{{\cal R}}_{\{1\}}\hat{{\cal R}}_{\{1\}}) - 1,
\end{align}
where $\hat{{\cal R}}_{\{i\}}$ are the rescaling factors of the spherical background
metric (see~\cite{Mewes:2020vic}), and we do not enforce the 
constraint~\eqref{trA_constraint} at all. Note that we relaxed the enforcement of the algebraic constraints only inside the horizon. 

\subsection{Filtering parallelization}

Here we describe the algorithm that we use to perform the FFT filtering across 
multiple compute nodes. To start, we note that there are two strategies to 
implement MPI-parallelized FFT
filtering: we could either use the parallel FFTW3 \cite{FFTW3} implementation; or gather data to 
be filtered, use serial FFTs to filter, and broadcast the filtered data back to 
their corresponding MPI ranks. We have chosen the second approach here, rather 
than relying on a particular parallel implementation of the FFTW3 library being 
installed on a given cluster. Note that even if we were to use the parallel 
implementation of FFTW3, we would still need to perform the MPI communicator split 
described below.

\begin{figure}
  \resizebox{0.9\columnwidth}{!}{
  \begin{tikzpicture}
    \tikzset{->-/.style={decoration={
    markings,
    mark=at position #1 with {\arrow{>}}},postaction={decorate}}}
    \pgfmathsetmacro{\xrad}{3}
    \pgfmathsetmacro{\xth}{2}
    \pgfmathsetmacro{\xph}{4}
    \draw [very thick, black] (0, 0) -- (0, \xph);
    \draw [very thick, black] (0, 0) -- (\xth, 0);
    \draw [very thick, black] (\xth, 0) -- (\xth, \xph);
    \draw [very thick, black] (0, \xph) -- (\xth, \xph);
    \draw [very thick, black] (\xrad+\xth, \xrad) -- (\xrad+\xth, \xrad+\xph);
    \draw [very thick, black] (\xrad+\xth, \xrad) -- (\xth, 0);
    \draw [very thick, black] (\xrad+\xth, \xrad+\xph) -- (\xth, \xph);
    \draw [very thick, black] (\xrad+\xth, \xrad+\xph) -- (\xrad, \xrad+\xph);
    \draw [very thick, black] (0, \xph) -- (\xrad, \xrad+\xph);
    \draw [very thick, black] (\xrad/2, \xph+\xrad/2+0.1) node [rotate=45, above] {$r$};
    \draw [very thick, black] (\xth/2, 0-0.15) node [rotate=0, below] {$\theta$};
    \draw [very thick, black] (-0.15, \xph/2) node [rotate=0, left] {$\varphi$};
    \foreach \i in {0.6, 1.2, ..., 2.4} {
        \draw [very thick, red] (\i, \xph+\i) -- (\xth+\i, \xph+\i);
        \draw [very thick, red] (\xth+\i, \i) -- (\xth+\i, \xph+\i);
    }
    \foreach \i in {0.3, 0.9, ..., 2.7} {
        \draw [very thick, red] (\xth/2+\i, \xph+\i) node [rotate=0] {\tiny{${\rm COMM}_r$}};
    }
    \pgfmathsetmacro{\xo}{6.5}
    \pgfmathsetmacro{\xrad}{0.9}
    \pgfmathsetmacro{\xth}{3}
    \pgfmathsetmacro{\xph}{6}
    \draw [very thick, black] (\xo, 0) -- (\xo, \xph);
    \draw [very thick, black] (\xo, 0) -- (\xo+\xth, 0);
    \draw [very thick, black] (\xo+\xth, 0) -- (\xo+\xth, \xph);
    \draw [very thick, black] (\xo, \xph) -- (\xo+\xth, \xph);
    \draw [very thick, black] (\xo+\xrad+\xth, \xrad) -- (\xo+\xrad+\xth, \xrad+\xph);
    \draw [very thick, black] (\xo+\xrad+\xth, \xrad) -- (\xo+\xth, 0);
    \draw [very thick, black] (\xo+\xrad+\xth, \xrad+\xph) -- (\xo+\xth, \xph);
    \draw [very thick, black] (\xo+\xrad+\xth, \xrad+\xph) -- (\xo+\xrad, \xrad+\xph);
    \draw [very thick, black] (\xo, \xph) -- (\xo+\xrad, \xrad+\xph);
    \draw [very thick, black] (\xo+\xth/2, 0-0.15) node [rotate=0, below] {$\theta$};
    \draw [very thick, black] (\xo+-0.15, \xph/2) node [rotate=0, left] {$\varphi$};
    \foreach \i in {1, ..., 2} {
        \draw [very thick, teal] (\xo+\i, 0) -- (\xo+\i, \xph);
        \draw [very thick, teal] (\xo+\i+\xrad, \xph+\xrad) -- (\xo+\i, \xph);
    }
    \foreach \i in {1, 2, ..., 3} {
        \draw [very thick, teal] (\xo+\i, \xph+\xrad/2) node [rotate=45] {\tiny{${\rm COMM}_{r \theta}$}};
    }
    \foreach \i in {1, ..., 5} {
        \draw [very thick, blue] (\xo, \i) -- (\xo+\xth, \i);
        \draw [very thick, blue] (\xo+\xth+\xrad, \i+\xrad) -- (\xo+\xth, \i);
    }
    \foreach \i in {1, 2, ..., 6} {
        \draw [very thick, blue] (\xo+\xth+0.5, \i) node [rotate=45] {\tiny{${\rm COMM}_{r \varphi}$}};
    }
    \node (a) at (2.5,0.6) {};
    \node (b) at (6.6,0) {};
    \draw [->-=0.5, very thick, black] (a) to [right] (b);
    \node (a) at (2.5+0.6,0.6+0.6) {};
    \node (b) at (6.6,0.9) {};
    \draw [->-=0.6, very thick, black] (a) to [] (b);
    \node (a) at (2.5,4+0.57) {};
    \node (b) at (6.6,6) {};
    \draw [->-=0.5, very thick, black] (a) to [right] (b);
    \node (a) at (2.5+0.6,4+0.57+0.6) {};
    \node (b) at (6.6+0.87,6+0.9) {};
    \draw [->-=0.5, very thick, black] (a) to [right] (b);
  \end{tikzpicture}
  }
  \caption{Overview of the different MPI communicator groups. The global communicator is split along the r direction, and then each r group is split in two different subgroups. For one subgroup, the second splitting is in the $\theta$ direction, while for the other, it is in the $\varphi$. }
  \label{fig:mpi_comm}
\end{figure}
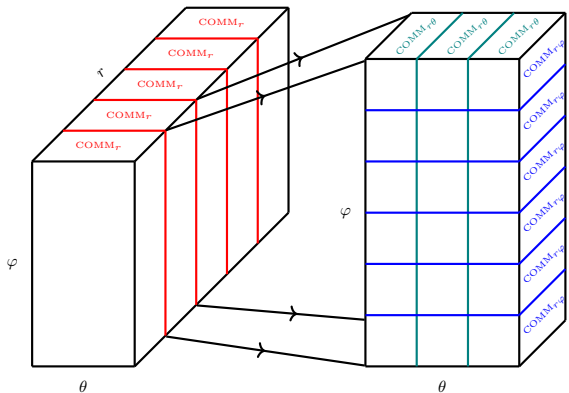

We first split the global MPI communicator in the $r$ direction into a set of smaller communicators : ${\rm COMM}_r$. In each group ${\rm COMM}_r$, all the member processes share the same $r$-coordinate range. Then we split each ${\rm COMM}_r$ further into two separate groups of communicators: ${\rm COMM}_{r \theta}$, which all share the same $r$ and $\theta$ ranges, and ${\rm COMM}_{r \varphi}$, which share the same $r$ and $\varphi$ ranges (see Fig.~\ref{fig:mpi_comm}).

The computational domain contained in ${\rm COMM}_{r \theta}$ covers $0 \le \varphi < 2\pi$. Because we need to extend the domain in $\theta$ to $0 < \theta < 2\pi$, we copy data from processes with $\varphi<\pi$ to the corresponding processes which owns $\varphi+\pi$ within the given communicator. At this point, processes in ${\rm COMM}_{r \theta}$ contain both the data corresponding to $0<\theta<\pi$ and the data corresponding to $\pi<\theta<2\pi$.

To filter in the $\varphi$ direction, we use MPI scatter / gather operations to redistribute the data so that each MPI process in ${\rm COMM}_{r \theta}$ contains a roughly equal number of arrays containing the full set of $\varphi$ points for some fixed values of $r$ and $\theta$. Each process then performs FFTs on these arrays (multiple FFTs at a time using 
{\sc OpenMP}), filters the transforms, and performs an inverse FFT. The filtered fields are then redistributed back to the original processes in the communicator. Filtering on $\theta$ proceeds in much the same way. However, since we use double covering, only half of the rings need to be transformed (i.e., one FFT --- filter --- inverse FFT operation actually filters two different $\theta$ {\it rings}), only half of the MPI processes need to perform the filtering (we could redistribute the data again so that all MPI processes can perform FFTs, but this would be less efficient due to the extra communication overhead).

\begin{table}
  \caption{The number of MPI processes in each direction and number of threads per MPI process used for the strong scaling test reported in Fig.~\ref{fig:scaling}.}\label{tab:MPI-distribution} 
  \begin{tabular}{llllc}
    \hline
    $N_{core}$ & $N_r$ & $N_\theta$ & $N_\varphi$ & Threads \\
    \hline
    \hline
    64 & 8 & 2 & 4 & 1 \\
    128 & 8 & 4 & 4 & 1 \\
    256 & 8 & 4 & 8 & 1 \\
    512 & 8 & 4 & 8 & 2 \\
    1024 & 16 & 4 & 8 & 2 \\
    2048 & 16 & 4 & 8 & 4 \\
    4096 & 16 & 8 & 16 & 2 \\
    \hline\hline 
  \end{tabular}
\end{table}

\begin{figure}
    \includegraphics{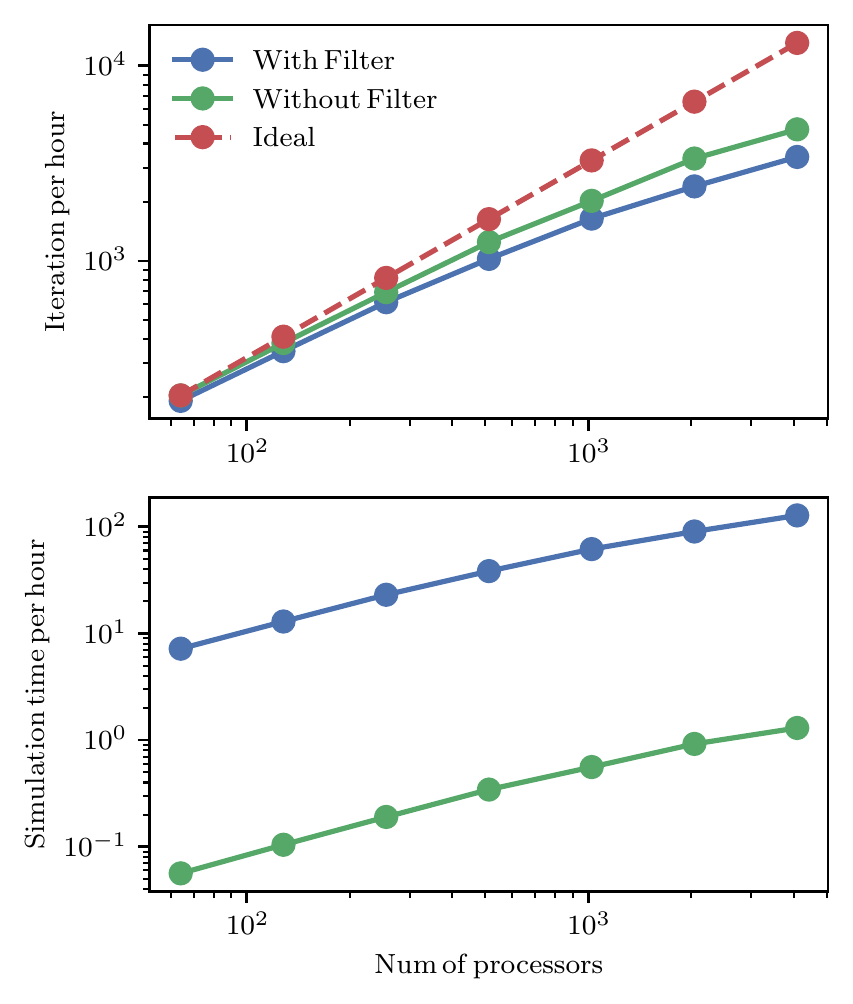}
    \caption{Strong scaling of \SphericalNR with the FFT filtering scheme. In the top plot, we show iterations per hour for both the code with and without the FFT filter. The code without the filter requires a time step 1/136 as small as with filtering. The bottom plot shows the actual performance in run time versus wall time.}
    \label{fig:scaling}
\end{figure}

In Fig.~\ref{fig:scaling}, we show the strong scaling performance of \SphericalNR with FFT filtering performed on the Frontera supercomputer at the Texas Advanced Computing Center. Here, we use a grid of $n_r=256, n_\theta=128, n_\varphi=256$ points and increase the number of cores from $64$ to $4096$. We consider both the performance of the parallel filter and unfiltered algorithm. The unfiltered algorithm requires a time step that is $\sim 136$ smaller than the filtered algorithm. When comparing just the number of iterations per unit wall time, we see that overhead of filtering is negligible up to about 300 cores. At 4096 cores, the nonfiltered algorithm is a factor of $1.38$ faster in terms of time steps
per unit wall time. Of course, when including the severe CFL limitations of the unfiltered algorithm (bottom panel of Fig.~\ref{fig:scaling}), we see that the filtered algorithm is actually roughly $100$ times faster (in terms of physical time) at 4096 cores.
In Table~\ref{tab:MPI-distribution}, we list the number of MPI in each direction and threads used for the scaling test.

\section{Results}
\label{sec:results}

\subsection{Vacuum spacetime test}

To evaluate the robustness of our new filtering algorithm, we simulate the same physical BH system previously used in~\cite{Mewes:2018szi} to introduce our new code. The system consists of a single spinning Bowen-York BH~\cite{Bowen:1980yu} (note that, unlike a Kerr BH, a spinning Bowen-York BH contains radiation due
to the initial data being conformally flat). We consider two physically equivalent scenarios: one in which the spin is aligned with the polar ($z$) axis, and another in which the spin is aligned with the $y$ axis. In~\cite{Mewes:2018szi}, we were able to show that when extracting the gravitational waves of the BH ringdown via the Weyl scalar $\Psi_4$, all modes up through $\ell=8$ were obtained with high accuracy for the aligned spin case. In these earlier results we used excision techniques to eliminate the severe CFL limitation of the origin in spherical coordinates. Here, we repeated those runs with filtering, rather than excision. While the two configurations are physically equivalent, they require significantly different numerical grid choices. For the case of the $z$ aligned spins, there is no azimuthal variation of the fields, whereas in the $y-$aligned case, a high azimuthal resolution is required. Consequently, in the $y$ case the CFL limitations associated with the polar axis are important. For reference, the CFL limitation for an excision run is 
$dt < r_{\rm exc} \sin(\theta_0/2) d\varphi$, where $r_{\rm exc}$ is the excision radius and $\theta_0$ is the $\theta$ grid point closest to the pole. 

The initial data were obtained using the {\sc TwoPunctures} code~\cite{2004PhRvD..70f4011A} (the mass, spin, and momentum parameters of one horizon were set to zero). The parameters associated with the data are a bare mass of $1M$ and a spin angular momentum of $0.8 M^2$. This corresponds to a BH with a horizon mass of $M_{\rm H} = 1.1811M$ and a dimensionless spin of $\chi = 0.5735$. We use the {\sc AHFinderDirect} 
thorn~\cite{2004CQGra..21..743T, Schnetter:2004mc} to find apparent horizons 
(AHs)~\cite{2007LRR....10....3T} and the
{\sc QuasiLocalMeasures} thorn~\cite{2003PhRvD..67b4018D,2006PhRvD..74b4028S} to calculate the angular momentum
of the apparent horizon during the evolution. The BH spin is measured using the 
flat space rotational Killing vector method~\cite{2007PhRvD..75f4030C} that was shown
to be equivalent to the Komar angular momentum~\cite{Komar:1958wp} in foliations adapted
to the axisymmetry of the spacetime~\cite{Mewes:2015vma}.

We denote the  simulations by Aligned(XX) and FFT(XX), where XX refers to the number of 
polar grid points, Aligned and FFT refer to simulations where the spins are aligned with the polar axis and nonaligned with the polar axis (hence both the $\theta$ and $\varphi$  FFT filters are needed). Note that the aligned cases were performed using excision of the BH interior (hence no filtering of any kind was used).

In Table~\ref{tab:grid}, we give the parameters for the computational grids used in all the simulations. For these runs, we set the filtering parameter ${\cal L}$ to ${\cal L}=2$.

\begin{table}
  \caption{The grid parameters for all simulations. Here $n_r$, $n_\theta$, and $n_\varphi$ are the number of grid points in the radial, polar, and azimuthal directions, respectively, and $dt$ is the timestep.}\label{tab:grid} 
  \begin{tabular}{lllll}
    \hline
    & $n_r$ & $n_\theta$ & $n_\varphi$ & $dt$ \\
    \hline
    \hline
    Aligned(64) & 2500 & 64 & 4  & $0.00490$ \\
    Aligned(96) & 3750 & 96 & 4  & $0.00327$ \\
    FFT(64) & 2500 & 64 & 128  & $0.001$ \\
    FFT(96) & 3750 & 96 & 192  & $0.00067$ \\
    \hline\hline 
  \end{tabular}
\end{table}

\begin{figure}
    \centering
    \includegraphics[width=.9\columnwidth]{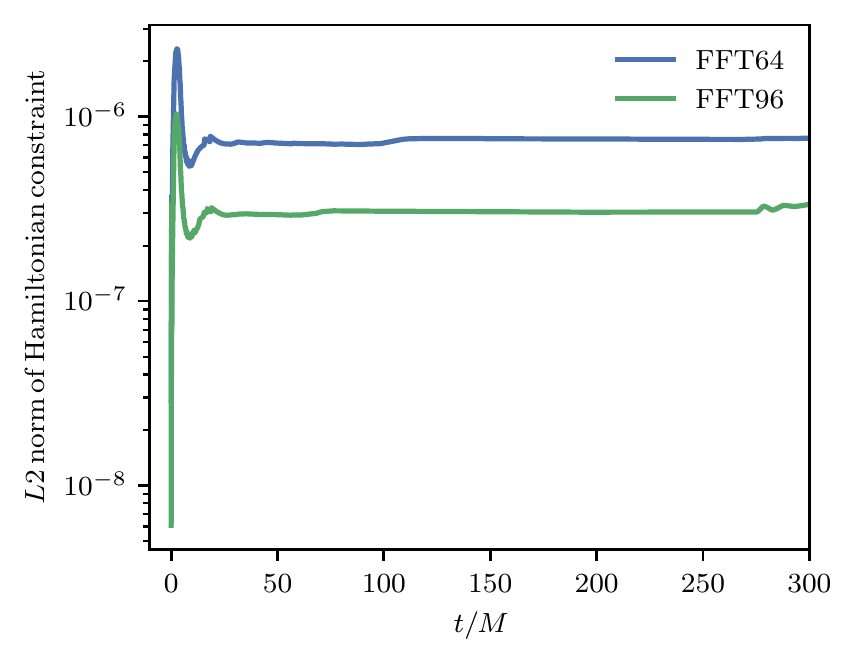}
    \caption{The $L^2$ norm of the Hamiltonian constraints versus time at two resolutions FFT64 and FFT96. Here, the norm only includes points outside the horizon. The drop in constraint magnitude between FFT64 and FFT96 is consistent with a convergence order of 2. }
    \label{fig:ham_BH}
\end{figure}

In Fig.~\ref{fig:ham_BH}, we show the $L^2$ norm of the Hamiltonian constraints for $n_\theta=64$ and $n_\theta=96$ for the using FFT filter case. After the initial oscillation, the constraint violation settles down to $7\times10^{-7}$ and $3\times10^{-7}$ for $n_\theta=64$ and $n_\theta=96$ respectively.

\begin{figure}
    \centering
    \includegraphics[width=.9\columnwidth]{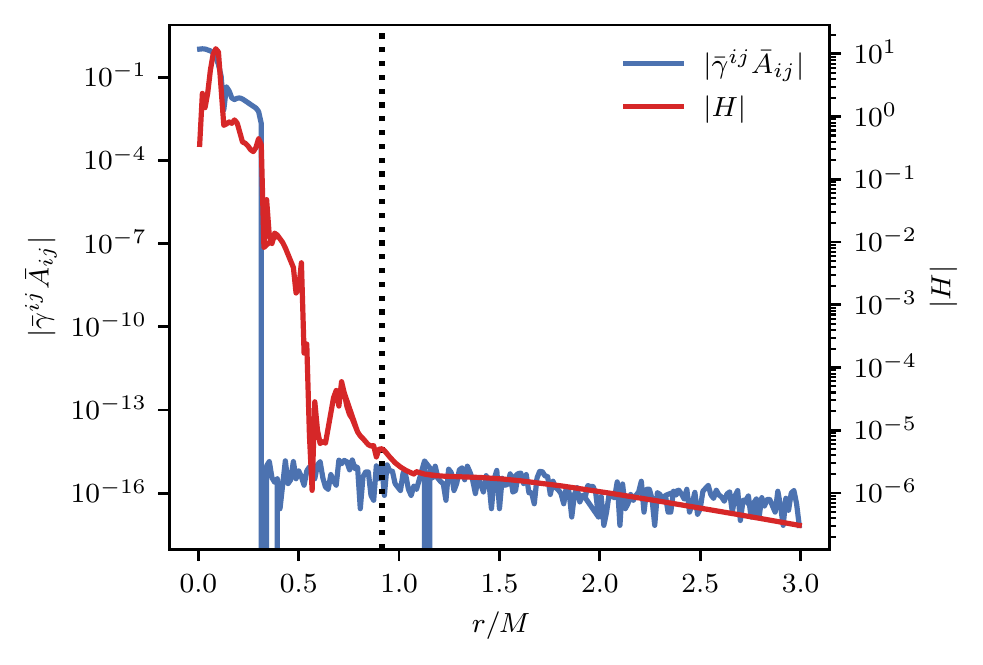}
    \caption{Algebraic constraints \eqref{trA_constraint} and Hamiltonian constraints along $\theta=d\theta/2, \varphi=0$ at $t=150M$ for case FFT96. The vertical line indicates the radius of apparent horizon.}
    \label{fig:algeC}
\end{figure}

In Fig.~\ref{fig:algeC}, we show the algebraic constraints violation \eqref{trA_constraint} and Hamiltonian constraints along $\theta=d\theta/2, \varphi=0$ at $t=150M$. The region where we do not enforce the algebraic constraints is within the horizon. All points on the horizon, and outside, have the algebraic constraints
enforced. As we can see, the algebraic constraint violations remain
below $10^{-13}$ even at radius of $r\sim0.35$ (here, the horizon radius is at $r=0.915$).
The algebraic constraint do increase closer to the puncture to about $1.0$. However,
this violation is nonpropagating.

\begin{figure}
  \includegraphics[width=.9\columnwidth]{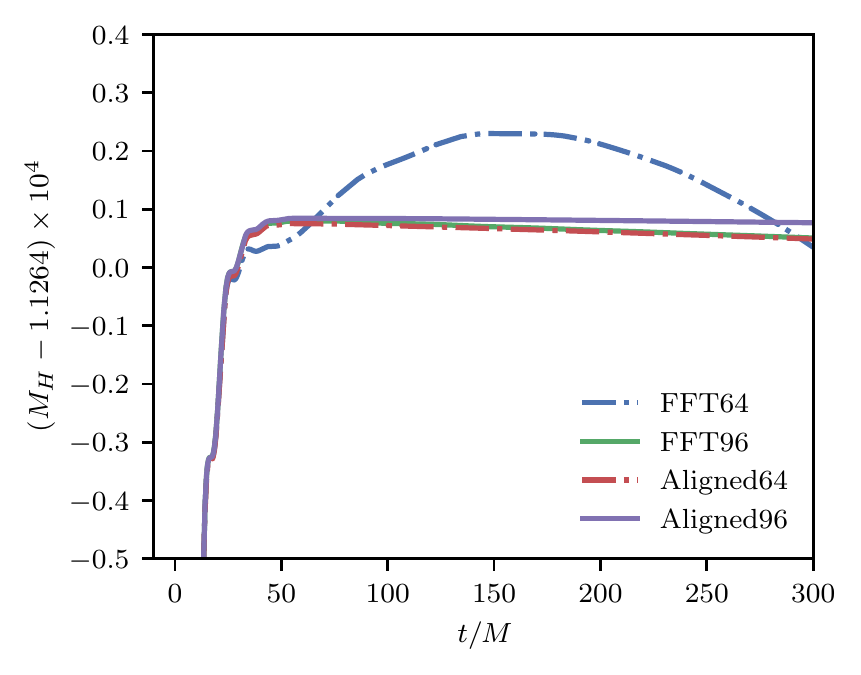}
  \caption{A comparison of the irreducible mass of the BH versus time as measured by {\sc AHFinderDirect} for $n_\theta=64$ and $n_\theta=96$ for the aligned case and using the double FFT filter. Here, mass conservation would imply that the irreducible mass settles to a constant value.} \label{fig:cmp_mass}
\end{figure}

\begin{figure}
    \includegraphics[width=.9\columnwidth]{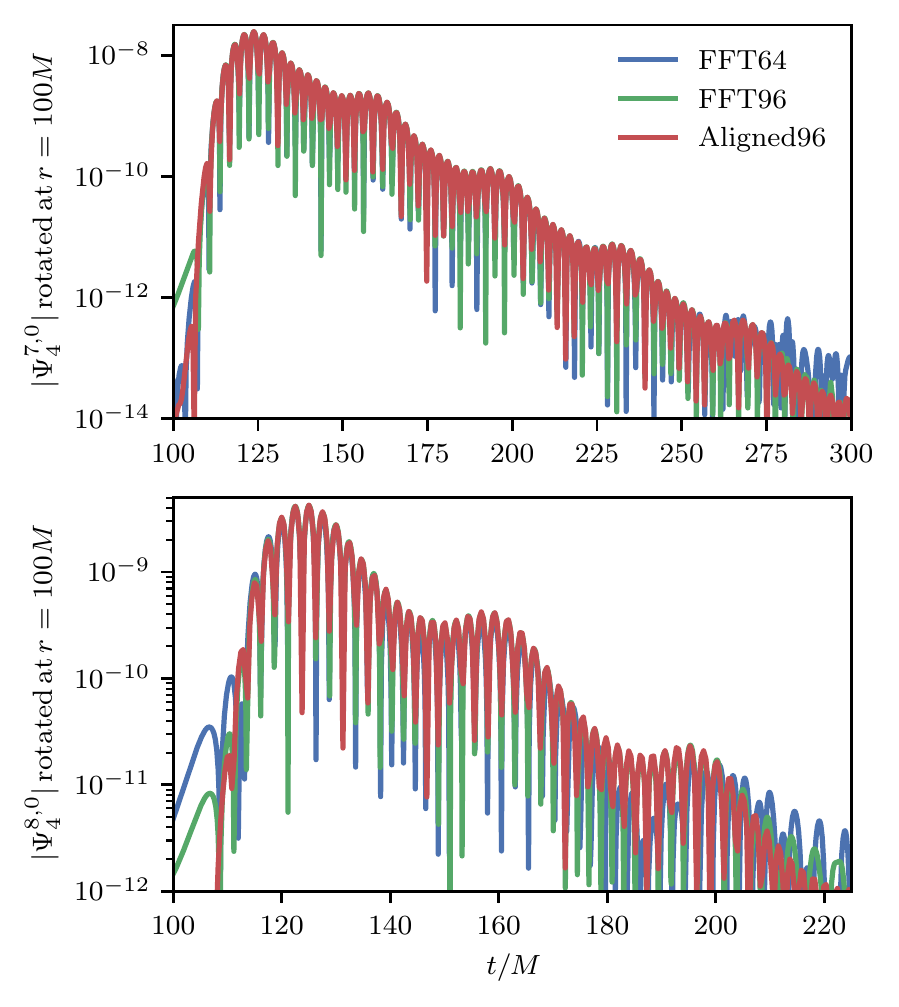}
    \caption{A comparison of the $\ell=7$ and $\ell=8$ modes of $\Psi_4$.}
    \label{fig:cmp_psi4}
\end{figure}

In Fig.~\ref{fig:cmp_mass}, we show the evolution of the irreducible mass $M_{\rm irr}$. After having absorbed some of the initial junk radiation, $M_{\rm irr}$ should remain constant absent of numerical errors (constraint violations occurring in free evolution can act as negative mass and result in an unphysical reduction of $M_{\rm irr}$~\cite{2011CQGra..28m4003M,2012PhRvD..86f4003R,2014PhRvD..89j4032O}). We see that the nonaligned case using the double FFT filter rapidly converges to the $M_{\rm irr}$ obtained in the aligned case, our reference point. In Fig.~\ref{fig:cmp_psi4} we show that the higher-order waveforms mode for the nonaligned case agree very well with the aligned case. Here, we rotate all waveforms of the nonaligned cases to a frame where the $z$ axis is aligned with the spin axis so that they should reproduce the aligned waveform. In this frame, only the $m=0$ modes are nontrivial in the continuum limit. The Aligned96 run is expected to have the smallest error, and we see that the double FFT filter runs rapidly converge to it. Note that differences are only apparent for the last few cycles.

\subsection{GRMHD tests}

\subsubsection{Off-center spherical explosion}
\begin{figure}
    \centering
    \includegraphics[width=\columnwidth]{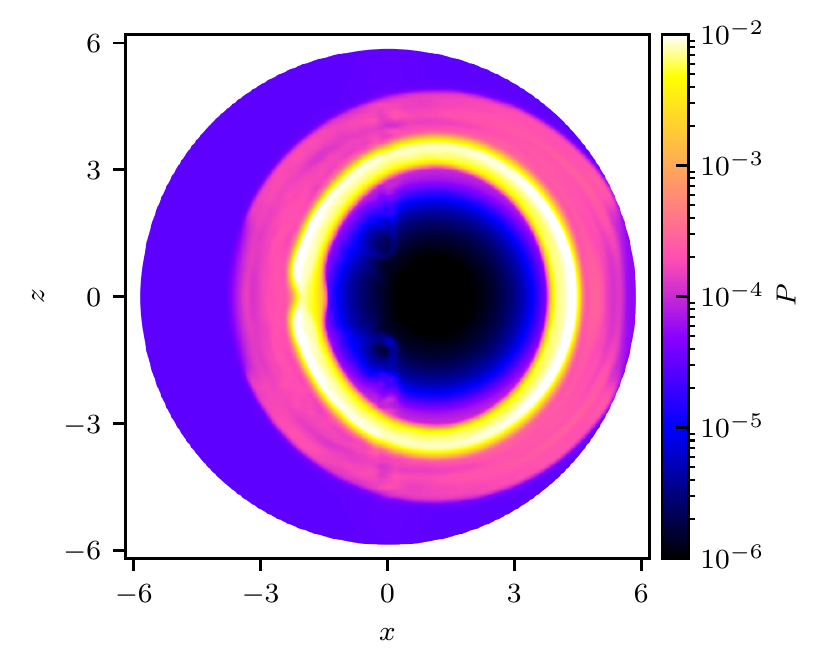}
    \includegraphics[width=\columnwidth]{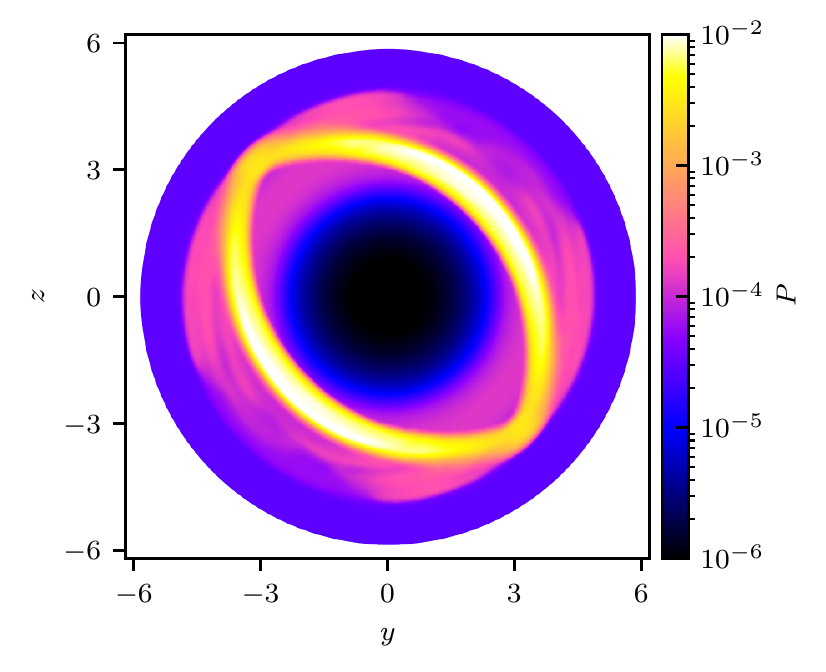}
    \caption{Pressure in the x-z and y-z plane at $t=4$, with initial magnetic field $B^z=0.1$ rotated by $45^\circ$ about $x$ axis.
    The top panel shows the shock front along the $xz$ plane ($y=0$), while the bottom shows the shock front on the plane $x=1.1$
    (note that the center of the explosion is at $x=1.1$). The bottom panel show the $x=1.1$ plane (the shock is
    centered along $x=1.1$).The flow morphology is clearly seen to be tilted by $45^{\circ}$.}
    \label{fig:tilted_pressure}
\end{figure}

To evaluate the effectiveness of our new filtering algorithm in solving challenging relativistic MHD problems, we selected the same test case as in our previous study~\cite{Mewes:2020vic}: a spherical explosion~\cite{CerdaDuran2008}, but with the explosion center intentionally displaced from the origin. 
The initial data consist of an overdense ($\rho=\num{1e-2}$, $p=1.0$) ball of radius 1.0.
From a radius of 0.8 outwards, the solution is matched in an exponential decay to the
surrounding medium ($\rho=1 \times 10^{-4}$, $p=3 \times 10^{-5}$). We use a
$\Gamma-$law equation of state (EOS) with $\Gamma=4/3$. The entire domain is initially
threaded by a constant magnitude magnetic field ($B^z=0.1$). The fluid three-velocity is 
set to zero everywhere in the domain initially. We use a fixed background Minkowski 
spacetime for this test problem. In order to remove any symmetries in the initial data and 
test the double FFT filtering in a full 3D setting, we offset to the center of the overdense 
region to $(x=1.1, y=0, z=0)$ and rotate the magnetic field by $45^{\circ}$ about the x
axis initially; the shock front of the explosion will therefore have to pass through the
origin and polar axis.

We use ($n_r=160, n_\theta=80, n_\varphi=160$) points, with the outer boundary $r_\text{max}=6.0$,
and use the double FFT filter to increase the time step from $dt=\num{7e-6}$ (stable limit without filtering) 
to $dt=\num{1e-3}$. Here we use the filtering parameter
$\mathcal{L}=4$.
We use MPWENO-Z9, as we found that the MP limiter helps the shock pass through origin and axis, 
tenth order finite differences in the curl of $A_i$, ninth order KO dissipation with a dissipation 
strength $\epsilon_{\rm diss} = 0.1$, global Lax-Friedrich fluxes with higher order flux corrections, and isotropic floors with 
$\rho_{\rm floor} = \num{1e-7}$, $(\rho \epsilon)_{\rm floor}=\num{1e-9}$.  

The final distribution at $t=4$ for the pressure $P$ (in the $x=1.1$ and $y=0$ planes) is shown in Fig.~\ref{fig:tilted_pressure}. Here we do see the shockfront propagating through the origin and poles, but there are small residual artifacts associated with them.  The test is particularly challenging here because radial flows near the origin need to be converted into angular flows around the origin (the lower radial face there would have size of zero). A similar complication arises on the poles when considering longitudinal flows. 
In this test, we see that while the magnetized shock largely passes through the origin and axes, there are visible artifacts. The shock front propagation through the origin is slightly delayed, leading to a bump on the shock front. There is also extra pressure in the vicinity of the pole which is the result of occasional primitive recovery failures, which is not unexpected, as the difficulty of this test lies in the primitive recovery~\cite{2011ApJS..193....6B}. We note that we  performed a similar test in a previous paper~\cite{Mewes:2020vic}, filtering only in the $\varphi$ direction (with a correspondingly smaller time step than used here) and used lower-order reconstruction methods. With the improvements to the robustness of our GRMHD code described in Sec.~\ref{sec:techniques} above, {\sc SphericalNR} is now able to evolve the off-center spherical explosion using higher order methods.

\subsubsection{Off-center neutron star}

Next, we turn to the dynamical spacetime evolution of NSs, testing the double 
FFT filter in the coupled spacetime and GRMHD evolution. Our first test is the evolution
of a stable rotating NS. We evolve model B2 of~\cite{StergioulasBU22004} and 
add a weak poloidal magnetic field initially. Similar to the spherical explosion, 
we place the center of the star off center at $(x,y,z)=(1,1,0)$ initially.
The fluid and spacetime initial data are generated with the {\sc RNS} 
code~\cite{StergioulasRNS1995}, which has been incorporated as the {\sc Hydro\_RNSID} thorn 
in the {\sc Einstein Toolkit}. 

Model B2 is differentially rotating and is described by a j-law profile
\begin{align}
    \Omega_c-\Omega=\frac{1}{\hat{A}^2R_e^2}
    \left[\frac{(\Omega-\omega)r^2\sin^2\theta e^{-2\nu}}{1-(\Omega-\omega)r^2\sin^2\theta e^{-2\nu}}\right]
\end{align}
where $R_e, \Omega_c$ are provided in Table~\ref{table:NS_models},
and $\hat{A}$ is a measure of the degree of differential rotation, which we set to $\hat{A}=1$.
After interpolating and coordinate transforming the fluid and spacetime data from {\sc Hydro\_RNSID} to the orthonormal basis in spherical coordinates, we add a weak poloidal magnetic field, following the vector-potential-based prescription of~\cite{Liu2008}:
\begin{align}
A_x&=-(y-y_c)A_b\left(1-\frac{\rho}{\rho_c}\right)^{n_s}\max(P_\text{cut}-P,0), \label{eq:ID_Ax}\\
A_y&= (x-x_c)A_b\left(1-\frac{\rho}{\rho_c}\right)^{n_s}\max(P_\text{cut}-P,0), \label{eq:ID_Ay}\\
A_z&= 0, \label{eq:ID_Az}\\
\varphi&= 0, \label{eq:ID_Phi}
\end{align}
where values of $A_b,\rho_c,n_s,$ and $P_\text{cut}$ are provided in 
Table~\ref{table:NS_models}. This choice of initial vector potential in Cartesian
coordinates results in a purely azimuthal vector potential and therefore, a purely 
poloidal magnetic field. $A_i$ is then transformed to the orthonormal basis in spherical 
coordinates and the initial magnetic is calculated from the curl of $A_i$.
While the EOS of the initial data are polytropic, we evolve the star with a $\Gamma$-law EOS 
with $\Gamma=2$. We use SSPRK54 for time integration, fourth order finite differences with fifth 
order KO dissipation with $\epsilon_{\mathrm{diss}} = 0.01$ in the spacetime evolution, the 
HLLE Riemann solver, WENO-Z9 reconstruction, tenth order finite difference in the curl of $A_i$, 
ninth order KO dissipation with $\epsilon_{\mathrm{diss}} = 0.01$, isotropic floors with 
$\rho_{\rm floor} = \num{5e-9}$, $(\rho \epsilon)_{\rm floor}=\num{5e-11}$, 
$W_{\mathrm{max}}=50$, and $\epsilon_{\mathrm{max}}=1$.
We evolved this off-centered NS configuration using four different resolutions: 
$n_r\times n_\theta \times n_\varphi =$
$256\times16\times32\;(h_0)$,
$512\times32\times 64\;(h_1)$,
$768\times 48\times 96\;(h_2)$,
and $1024\times 64 \times 128\;(h_3)$.

Since the NS is not centered on the origin, truncation errors introduce asymmetries into its evolution. Consequently, the NS drifts from its starting position (where it would remain 
if the grid was adapted to the symmetries of the star). Based on the results 
from our previous test, matter flows through the origin are impeded relative to flows across 
nonsingular points. Therefore, it is not apparent a priori that the drift will consistently 
converge to zero with increased resolution.

\begin{figure}
    \includegraphics[width=.9\columnwidth]{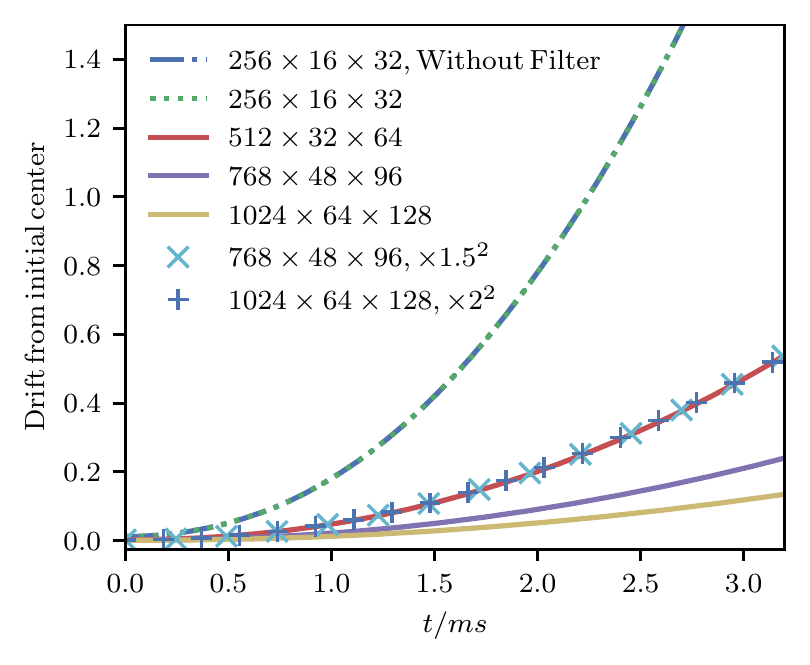}
    \caption{Drift of the NS center of mass for different resolutions. The plot shows relative drift of the center of the NS for four different resolutions
    : $h_0\,(256\times16\times32)$, $h_1\,(512\times32\times64)$, $h_2\,(768\times48\times96)$, $h_3\,(1024\times64\times128)$. The crosses show the relative drift of the $h_2$ run after rescaling by the square of the ratio of the $h_2$ to $h_1$ resolution grid sizes. The pluses show the same rescaled drift, but for the $h_3$ case. These rescaled drifts show clear second-order convergence.
    The dashed and doted curve are two $h_0$ resolution runs with
    and without filtering. They are on top of each other showing that the filter
    has no effect on the drift of the NS.
    }
    \label{fig:drift}
\end{figure}

In Fig.~\ref{fig:drift} we show that the drift of the NS from its initial location converges to zero to second order.
In the figure, the cross and plus symbols show the drift of the $h_2$ and $h_3$ resolutions overlap with the $h_1$ resolution drift if the former two are multiplied by the square of the ratio of their resolutions to the $h_1$ resolution. (Note that although we use higher-order reconstruction, the main algorithm is still second-order convergent).
This indicates that the errors associated with 
flows through the singular regions do converge away.

\begin{figure}
    \includegraphics[width=.9\columnwidth]{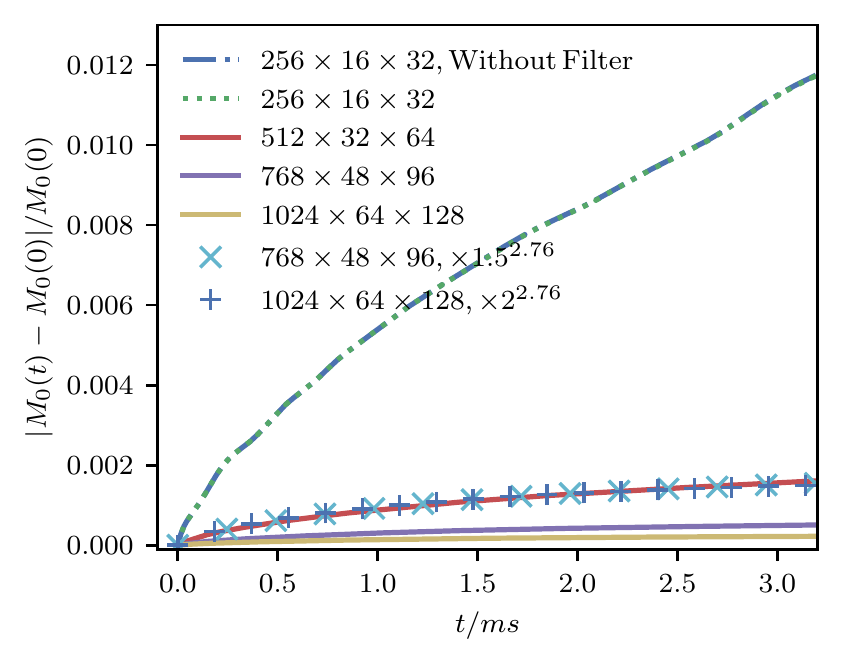}
    \caption{Conservation of total rest mass for different resolutions. The plot shows relative change in the total mass of the NS for four different resolutions: $h_0\,(256\times16\times32)$, $h_1\,(512\times32\times64)$, $h_2\,(768\times48\times96)$, $h_3\,(1024\times64\times128)$. The crosses show the relative change in mass  of the $h_2$ resolution run after rescaling by the ratio of the $h_2$ to $h_1$ resolution grid sizes to the power 2.76. The pluses show the same rescaled mass change, but for the $h_3$ case. Hence the convergence order is between 2 and 3.
    The dashed and doted curve are two $h_0$ resolution runs with
    and without filtering. They are on top of each other showing that the filter
    has no effect on the conservation of total rest mass.
    }
    \label{fig:rest_mass}
\end{figure}

Figure.~\ref{fig:rest_mass} shows the relative error in the conservation of total rest mass 
in the domain. As explained in ~\cite{Mewes:2020vic}, our implementation of the continuity 
equation in the reference metric formalism does not conserve total rest mass to round-off, but the total mass loss/gain will converge to zero with increasing resolution (additionally, some of the fixes in the primitive recovery and the need to use an artificial atmosphere will also break conservation of total rest mass). The plot shows several points on the $h_2$ and $h_3$ resolution curves after 
multiplying by the inverse of the ratio of the $h_2$ / $h_3$ resolution with the $h_1$ 
resolution raised to the power of 2.76 (i.e., demonstrating between second and third-order 
convergence). 

\begin{figure}
    \includegraphics[width=.9\columnwidth]{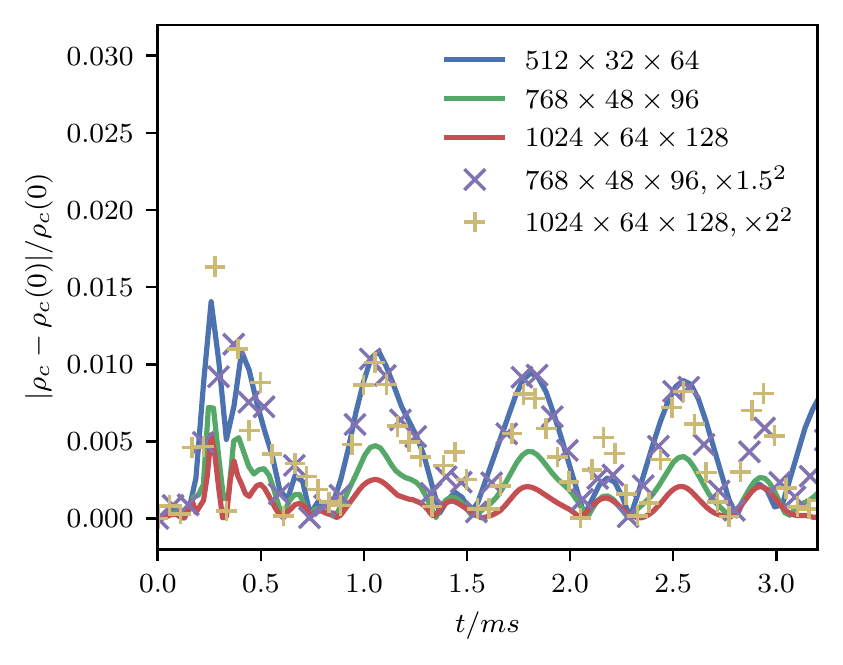}
    \caption{Maximum density versus time for different resolutions. The plot shows relative change in the central density of the NS versus time for three different resolutions. The crosses show the relative change of the $h_2$ resolution run after rescaling by the square of the ratio of the $h_2$ to $h_1$ resolution grid sizes. The pluses show the same rescaled change, but for the $h_3$ resolution case. In regions where the relative change in the central density is large, second-order convergence is observed.}
    \label{fig:central_density}
\end{figure}

The evolution of the fractional change in the maximum density is shown in 
Fig.~\ref{fig:central_density}. Truncation errors in the spacetime evolution, the interface of the NS surface and atmosphere as well as the asymmetric grid in our setup 
introduce perturbations of the NS that cause the central density to oscillate. As model B2 is
stable, these should converge away with grid resolution in the absence of added perturbations.
The convergence of the oscillations of the central density is less clean than either the drift or the total mass, likely due to the $h_1$ resolution case 
being too low. Here, we see that the oscillations generally converge away at second-order when the perturbations are large. 

\begin{figure}
    \includegraphics[width=.9\columnwidth]{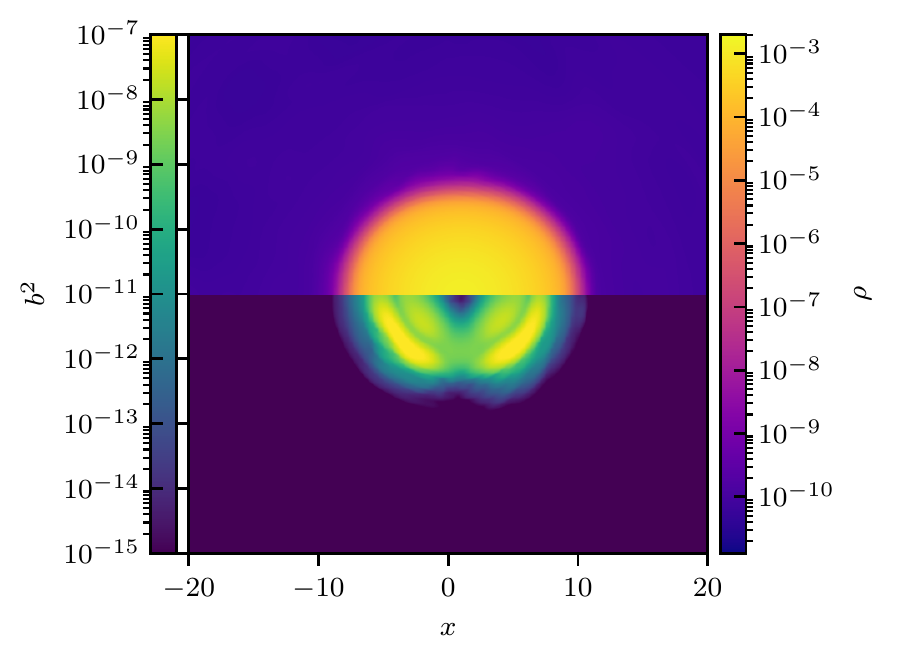}
    \caption{$\rho$ and $b^{2}$ evolution in $x-z$ plane of $y=1$ of model B2 at $t=3$ ms.}
    \label{fig:offcenteredns_rho_bsq_xz}
\end{figure}

Finally, Fig.~\ref{fig:offcenteredns_rho_bsq_xz} shows the distributions of $\rho$ and 
$b^2$ in the $y=1$ plane at $t=3{\rm ms}$. The star remains stable
and the stellar surface is well captured by the code.
During the evolution, the quantity $b^2$ develops a richer morphology than it had at the beginning. This test shows that the double FFT filter method works in a setup not adapted
to the symmetries of the coordinate system in a dynamical spacetime and GRMHD evolution.

\begin{table}
\caption{Main properties of the relativistic polytrope models B2~\cite{StergioulasBU22004} and U11~\cite{2013PhRvD..88j4028F}:
central rest-mass density $\rho_{c}$,
rest- and gravitational masses $M_0$ and $M$,
the dimensionless angular momentum $J/M^2$,
the proper equatorial radius $R_e$,
the angular velocities at the axis $\Omega_c$ and at the equator $\Omega_e$,
the ratio of polar and equatorial radii of the star $r_p/r_e$,
the ratio of kinetic energy and gravitational binding energy $T/|W|$,
the adiabatic index $\Gamma$,
the polytropic constant $K$,
and the constants prescribing the initial magnetic field $A_b$, $n_s$ and $P_{\mathrm{cut}}$ (see main text for details).}
\begin{tabularx}{\columnwidth}{ZZZ}
\toprule
\noalign{\vskip 1mm}
& B2 &U11 \\
\hline
\noalign{\vskip 1mm}
$\rho_{c}$  &\num{1.28e-3} &\num{1.092e-4} \\
$M_0$       &1.592 &1.508 \\
$M$         &1.478 &1.462 \\
$J/M^2$     &\num{3.177e-1} &1.660 \\
$R_e$       &9.92 & 23.3\\
$\Omega_c$  &\num{1.53e-2} &\num{1.29e-2} \\
$\Omega_e$  &\num{9.45e-3} &\num{8.61e-3} \\
$r_p/r_e$   &0.900 & 0.250\\
$T/|W|$     &\num{2.574e-2} &\num{2.743e-1} \\
$\Gamma$    &2  & 2\\
$K$         &100  & 100\\
$A_b$       &10 & 10\\
$n_s$       &3 & 2\\
$P_{\mathrm{cut}}$  &\num{6.542e-6} & \num{9.083e-8}\\
\noalign{\vskip 1mm}
\botrule
\end{tabularx}
\label{table:NS_models}
\end{table}
\begin{figure*}[!h]
  \begin{tabular}{cc}
    \subfloat[$t=9~\mathrm{ms}$.]{
      \includegraphics[width=0.45\textwidth,clip]{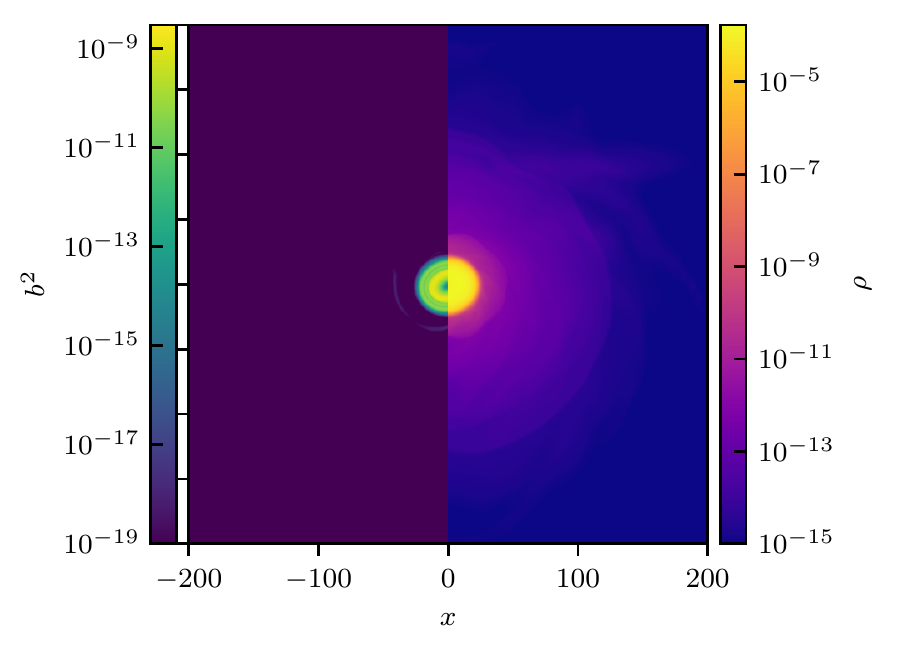}
      \label{fig:barmode_rho_xy_09}
    }
    &
    \subfloat[$t=9~\mathrm{ms}$.]{
      \includegraphics[width=0.45\textwidth,clip]{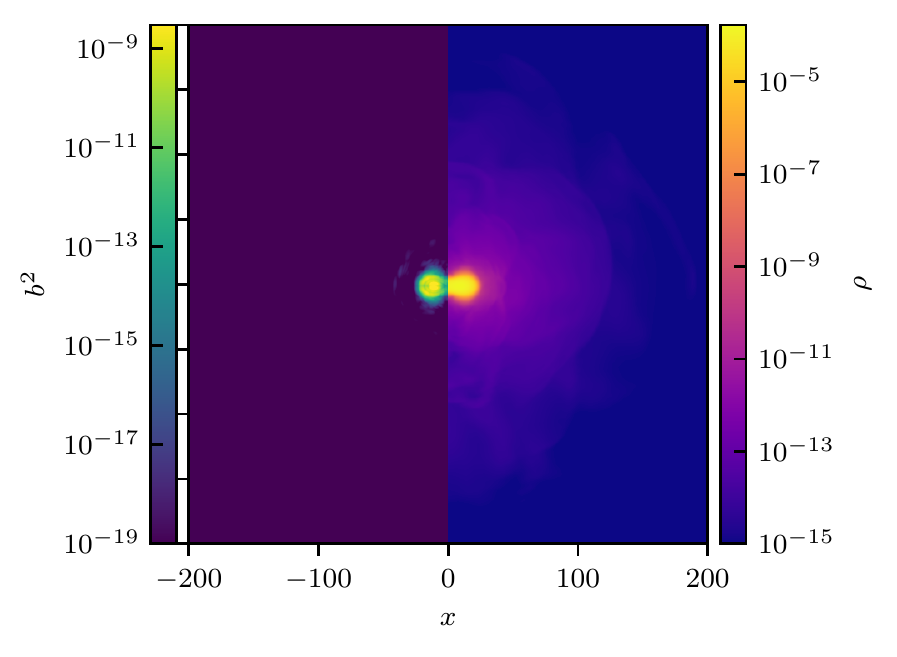}
      \label{fig:barmode_bsq_xy_09}
    } \\
    \subfloat[$t=18~\mathrm{ms}$.]{
      \includegraphics[width=0.45\textwidth,clip]{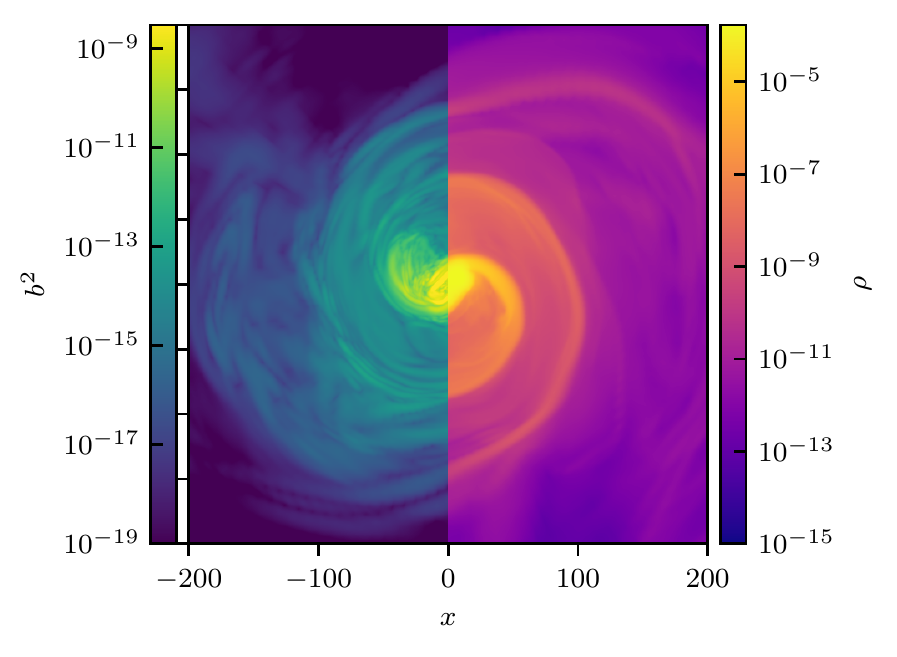}
      \label{fig:barmode_rho_xy_18}
    }
    &
    \subfloat[$t=18~\mathrm{ms}$.]{
      \includegraphics[width=0.45\textwidth,clip]{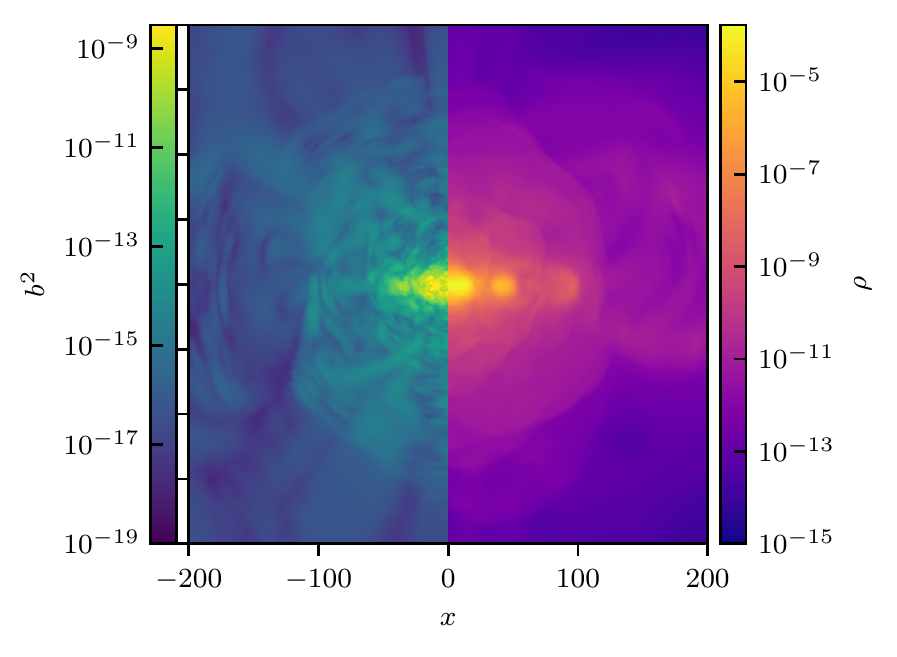}
      \label{fig:barmode_bsq_xy_18}
    } \\
    \subfloat[$t=45~\mathrm{ms}$.]{
      \includegraphics[width=0.45\textwidth,clip]{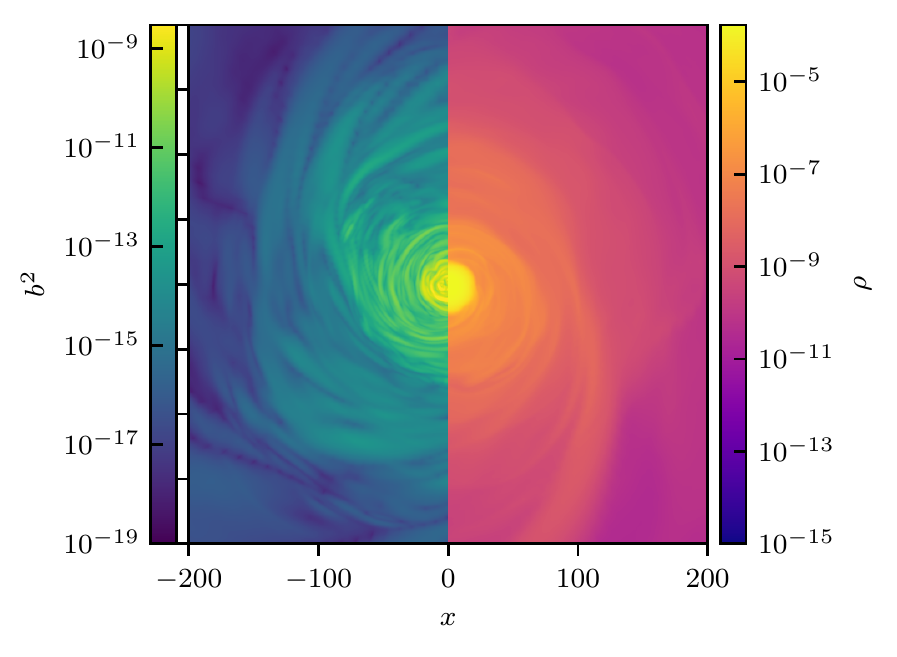}
      \label{fig:barmode_rho_xy_45}
    }
    &
    \subfloat[$t=45~\mathrm{ms}$.]{
      \includegraphics[width=0.45\textwidth,clip]{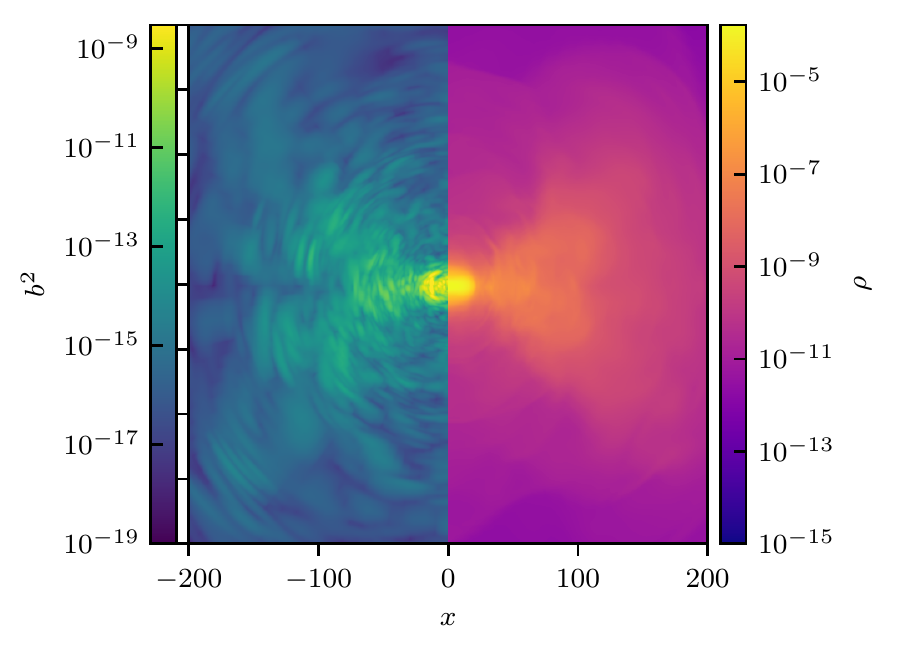}
      \label{fig:barmode_bsq_xy_45}
    }
  \end{tabular}
  \caption{Snapshots of $b^2$ and $\rho$ on $x-y$ (left column) and $x-z$ (right column) plane of model U11 at $t=9$ms (top row), $t=18$ms (middle row) and $t=45$ms (bottom row).}
  \label{fig:barmode_rho_bsq_xy}
\end{figure*}

\subsubsection{Dynamical bar-mode instability}

Our final code test serves as a proxy for the postmerger remnant of a binary NS merger: the 
evolution of the dynamical bar-mode instability of a NS in a dynamical 
spacetime~\cite{2000ApJ...542..453S} (see also the reviews 
in~\cite{2003CQGra..20R.105A,2017LRR....20....7P}). We evolve the dynamical bar-mode instability
in model U11 from~\cite{2013PhRvD..88j4028F}. The initial data configuration is outlined in Table~\ref{table:NS_models}. To generate the initial NS, we use again the {\sc Hydro\_RNSID} thorn and place the star at the 
origin and such that its spin axis is aligned with the polar axis. To trigger the growth of
the bar-mode instability, we perturb the pressure by 5\% with random noise initially. We add an 
initial poloidal magnetic field determined by Eqs.~\eqref{eq:ID_Ax}-\eqref{eq:ID_Phi},  setting 
$x_c=y_c=0$ since the star is initially centered on the coordinate origin. The constants $A_b, n_s,$ and $P_{cut}$ are chosen in such a way that $b^2_{\rm max}(t=0) \approx \num{1e15} G$,
which is classified as moderate field strength in~\cite{2013PhRvD..88j4028F}, where it was shown
that such a magnetic field is not strong enough to suppress the development of the bar-mode
instability.

We use SSPRK54 for time integration, 
fourth order finite differences with fifth order KO dissipation with $\epsilon_{\mathrm{diss}} = 0.05$ in the spacetime evolution, the HLLE Riemann solver,  
WENO-Z9 reconstruction, tenth order finite difference in the curl of $A_i$, ninth order KO 
dissipation with $\epsilon_{\mathrm{diss}} = 0.001$, $W_{\mathrm{max}}=50$, 
$\epsilon_{\mathrm{max}}=1$, and radially dependent floors with 
$\rho_{\rm floor} = \num{1e-19}$, $(\rho \epsilon)_{\rm floor}=\num{1e-24}$, and 
$r_{\mathrm{min}}=20$. The improvements made to our primitive recovery scheme described in Sec.~\ref{sec:techniques} allow us to use these very low floor values and still stably 
evolve the magnetic field everywhere in the domain.

\begin{figure}
    \includegraphics[width=.9\columnwidth]{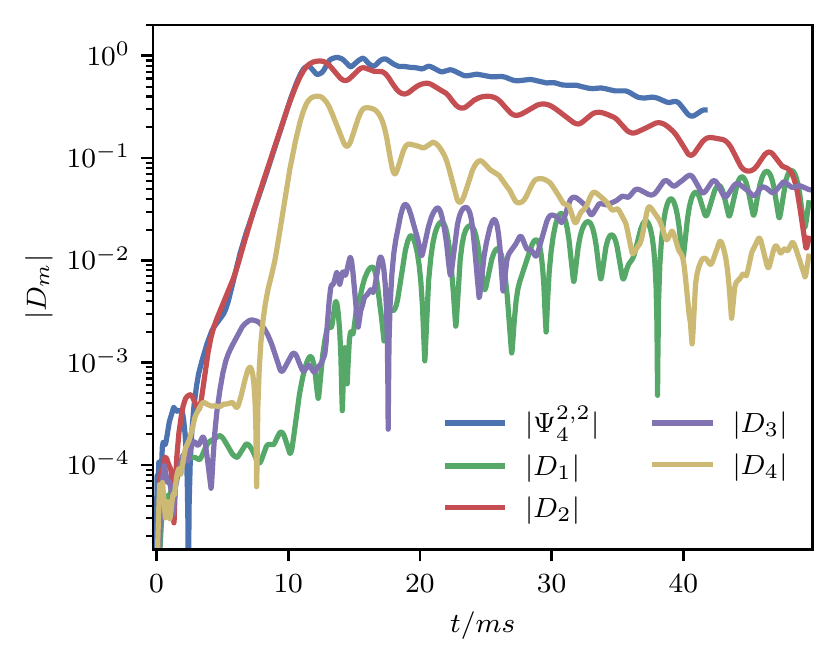}
    \caption{
    Evolution of the first four nonaxisymmetirc matter distribution modes and the $(\ell=2,m=2)$ mode of $\Psi_4$. As expected, the $m=2$ mode of the density distribution is the dominant contribution to the waveform quadrupole mode.
    }
    \label{fig:fourier_m2_vs_psi2_l2m2}
\end{figure}

In Fig.~\ref{fig:barmode_rho_bsq_xy}, we show snapshots of both $b^2$ and $\rho$ in the $xy$ and $xz$ planes at select times: prior to the onset of the bar-mode instability, the fully 
developed bar-mode including spiral arms, and at late times when the NS has become nearly 
axisymmetric again and is now surrounded by a disk with a turbulent magnetic field as a 
result of the growth of the magnetorotational 
instability~\cite{velikhov1959stability,1960PNAS...46..253C,1991ApJ...376..214B,1998RvMP...70....1B}. To quantify the development and subsequent saturations of the bar-mode instability, we plot the time evolution of the azimuthal Fourier modes of $\rho$,
\begin{align}
    D_m=\int\alpha\sqrt{\gamma}\rho e^{-i m \varphi}d^3x
\end{align}
in Fig.~\ref{fig:fourier_m2_vs_psi2_l2m2}. The evolution is very similar to the results 
of~\cite{2007PhRvD..75d4023B,2013PhRvD..88j4028F} (see in particular the schematic of mode evolution shown in Fig. 8 in~\cite{2007PhRvD..75d4023B}). As observed in~\cite{2013PhRvD..88j4028F}, the chosen initial magnetic field is not strong enough to disrupt the dynamical bar-mode instability. 

To test the correctness of the coupled spacetime and fluid evolution, we also plot the $(\ell=2, m=2)$ mode of the Weyl scalar $\Psi_4$ in Fig.~\ref{fig:fourier_m2_vs_psi2_l2m2} to compare it compare it with $D_2$. The idea here is that $\ell=2, m=2$ quadrupole moment of the gravitational waveform should be dominated by the azimuthal $m=2$ mode of the density distribution. This, in turn, should be the dominant contribution to the gravitational wave signal. As seen in 
Fig.~\ref{fig:fourier_m2_vs_psi2_l2m2}, the two modes are strongly correlated (after accounting for a time
translation of $t=1600$ due to the distance to where we extract $\Psi_4$, as well as an arbitrary factor to make it easier to see that both curves have the same growth rate). 

\begin{figure}
    \includegraphics[width=.9\columnwidth]{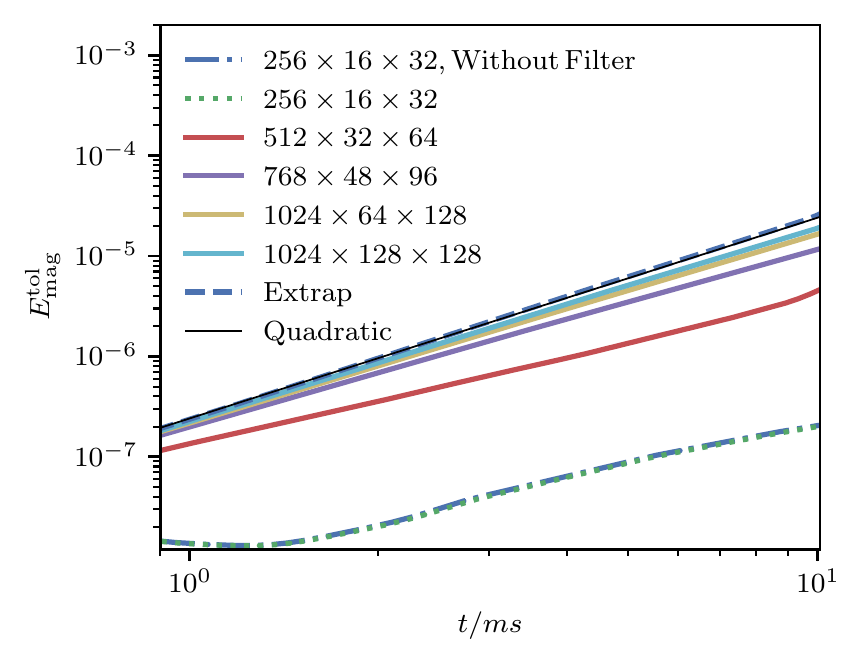}
    \caption{The toroidal component of the magnetic energy $E^{tor}_{mag}$ for early times and different resolutions. Prior to the formation of the instability, the magnitude of the $E^{tor}_{mag}$ is expected to increase quadratically  in time~\cite{2006PhRvD..73j4015D,2013PhRvD..88j4028F} (i.e., as $t^2$). The plot shows $E^{tor}_{mag}$ for different resolutions and a Richardson extrapolations assuming second-order convergence, as well as quadratic curve $y=c t^2$ (chosen to match the initial behavior of the Richardson extrapolated curve). 
    The lowest resolution run ($256\times16\times32$) was run with and without the filter. Since there is no visible difference in the toroidal magnetic energy growth when the filter is added or removed, this demonstrates that truncation error dominates over any errors introduced by the filter.
    }
    \label{fig:toroidal_mag_energy}
\end{figure}

In Fig.~\ref{fig:toroidal_mag_energy}, we plot the \enquote{toroidal} component of the magnetic energy~\cite{2013PhRvD..88j4028F},
\begin{align}
    E_{\text{mag}}^{\text{tor}}=\int \sqrt{\gamma}\frac{1}{2}B_{\parallel}B_{\parallel} d^3x,
\end{align}
where $B_{\parallel}=B^jv_j/(v^iv_i)^{1/2}$ is the \enquote{parallel} part of the magnetic 
field along the direction of the fluid motion. Note that $v^i$ is the Valencia 3-velocity and the indices are raised and lowered with the spatial metric $\gamma_{ij}$.
At early times, when the bar-mode has not yet developed, the differential rotation 
profile of NS winds up the poloidal magnetic field; therefore, the toroidal 
magnetic energy is expected to grow quadratically with time (i.e., 
$\sqrt{E_{\text{mag}}^{\text{tor}}} 
\propto t$)~\cite{2006PhRvD..73j4015D,2013PhRvD..88j4028F}. Here, we performed runs at four 
different resolutions
and measured the growth of this energy. We find that the toroidal component of the energy 
converges between first and second order. We plot  $E_{\text{mag}}^{\text{tor}}$ for the four 
resolutions and a Richardson extrapolation of these results in 
Fig.~\ref{fig:toroidal_mag_energy}. Prior to the onset of the bar-mode (i.e., for $t\lesssim 10 
{\rm ms}$), the Richardson extrapolated $E_{\text{mag}}^{\text{tor}}$ grows as $t^2$ (i.e., 
has a slope of $2$ on a log-log plot), with the lower resolution simulations exhibiting 
successively smaller exponents, showing that the code converges to the expected behavior.
The lowest resolution run ($256\times16\times32$) was run with and without the filter. Since there is no visible difference in the toroidal magnetic energy growth when the filter is added or removed, this indicates that truncation error dominates over any errors introduced by the filter.

\begin{figure}
    \includegraphics[width=.9\columnwidth]{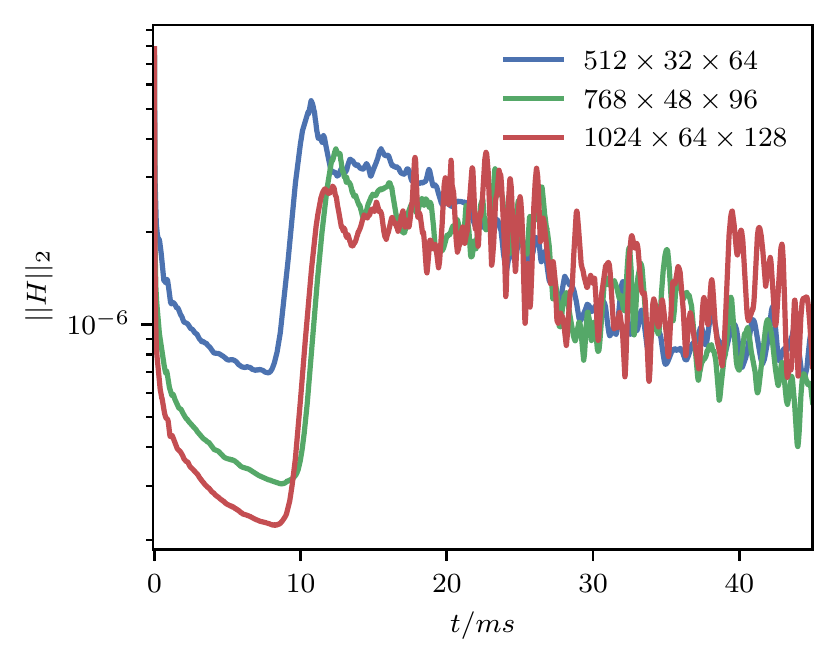}
    \caption{The $L^2$ norm of the Hamiltonian constraint for the bar-mode-unstable star. Prior to the onset of the bar-mode instability, we see convergence (to zero) at $~1.5$ order.  Note that, after the onset of the instability (which is seeded by random perturbations), the constraint violations remain small and do not exhibit exponential growth.}
    \label{fig:hamiltonian_norm2}
\end{figure}

In Fig.~\ref{fig:hamiltonian_norm2}, we show the $L^2$ norm of the Hamiltonian constraint for the three resolutions. Note that the onset of the bar-mode instability is triggered by random perturbations, so the time of this blowup is not expected to converge. Prior to the onset, we do see convergence of the constraints (except at $t=0$, where the constraint violations due to the initial perturbations dominate). Note that the constraint violations, after the initial bar-mode instability starts, remain roughly constant (with a small damping in time due to the 
constraint damping of the fCCZ4 spacetime evolution system).

\section{Discussion}
\label{sec:discussion}
Spherical coordinates are an attractive choice for GRMHD simulations of systems that possess (approximate) spherical- or axisymmetry, in particular they conserve fluid angular momentum and allow for a lower number of grid points compared to Cartesian coordinates. However, solving hyperbolic partial differential equations in spherical coordinates with high resolution suffers from the well-known CFL limitation resulting in prohibitively small time steps. In this work, we have developed a double FFT filtering algorithm for 
the coupled evolution of dynamical spacetimes and GRMHD in the {\sc SphericalNR}/\ET framework to ameliorate these CFL limitations by filtering both great circles in $\theta$ and coordinate rings in $\varphi$ depending on radius and latitude. Smooth fields are filtered by exponentially dampening CFL-unstable modes, while a hybrid exponential/Gaussian filter is used to filter fluid fields that can become discontinuous, in order to avoid Gibbs phenomenon when filtering. For these fields, the filter switches automatically between exponential and Gaussian filtering using Jameson's shock detector. 

The double FFT filter presented here is fully {\sc MPI}-parallelized, allowing for domain decomposition in the angular coordinates even though the FFT filter is a global operation (per great circle or azimuthal coordinate ring). Importantly, we showed good strong scaling properties of the algorithm up to thousands of cores. With the new filtering algorithm, we increase the time step from $dt \sim dr/2 \sin\left(\pi/(2 n_{\theta}) \right) 2 \pi/n_{\varphi}$ to a time step using effective
$n_{\theta}<10$ and $n_{\varphi}<10$, which can result in a time step orders of magnitude larger when filtering in high angular resolution simulations.

We have performed extensive testing of the new {\sc SphericalNR} code with the double FFT filter both in vacuum spacetimes and spacetimes coupled to GRMHD to evaluate its robustness and effectiveness. 
We performed code tests in which we deliberately subject the {\sc SphericalNR} with filtering to situations not adapted to the spherical coordinate system: the evolution of a Bowen-York BH with its spin axis aligned with the $y$ axis, and off-center simulations of a magnetized spherical explosion and a stable magnetized NS. These tests require significantly more angular resolution in the $\varphi$ coordinate than their counterparts when adapted to the symmetries (all three problems are axisymmetric) to produce accurate results.
For the vacuum test, we repeated the vacuum spacetime test of the original implementation of {\sc SphericalNR}~\cite{Mewes:2018szi}, showing that the FFT filtering is able to quickly achieve convergence to the expected results for spinning Bowen-York BHs, comparing to the axisymmetric result when the spin is aligned with the $z$ axis. Next, we successfully tested our new implementation by performing two challenging off-center GRMHD simulations: an off-center spherical explosion and the evolution of an off-center magnetized stable NS. Our code was able accurately reproduce known results and shown the expected convergence.

Finally, we evolved a magnetized NS model that is unstable against the dynamical bar-mode instability. This test was chosen as the dynamical bar-mode formation and saturation mimics the late stages of a BNS merger forming spiral arms and settling to a stable NS surrounded by a hot accretion disk. This test showed convergence and the correct coupled evolution of the dynamical spacetime and matter by showing a clear correlation between the growth rate of the $m=2$ azimuthal Fourier mode in the density distribution and the $\ell=2,m=2$ mode of the Weyl scalar $\Psi_4$. This type of test is important to show that {\sc SphericalNR} is capable of very long-term simulations of the postmerger remnant of BNS such as hypermassive NS and BH accretion disks. With our new FFT filtering algorithm these simulations can be performed in spherical coordinates, while maintaining convergent behavior.

Avoiding the CFL time step limitation with the double FFT filter presented in this work, {\sc SphericalNR} is well suited to simulate long-term BNS post merger remnants. While previous studies of BNS post mergers with spherical codes~\cite{2022PhRvD.106h3015L} used a fixed metric approach, our framework includes a fully dynamical metric. This will allow us to study the lifetime of hypermassive NS remnants and jet formation, and other interesting astrophysical scenarios, including gravitational core collapse and accretion into single or binary black hole systems.

\begin{acknowledgments}

The authors would like to thank Thomas W. Baumgarte, Pablo Cerd{\'a}-Dur{\'a}n, Luciano Combi, Eirik Endeve, Roland Haas, J. Austin Harris, W. Raphael Hix, Jay V. Kalinani, Eric Lentz, Carlos O. Lousto, Jens F. Mahlmann, O. E. Bronson Messer, Scott C. Noble, Martin Obergaulinger, David Radice, and Erik Schnetter for useful discussions. We gratefully acknowledge the National Science Foundation (NSF) for financial support from Grants No. \ PHY-2110338, No.\ OAC-2004044/1550436, No.\ AST-2009330, No.\ OAC-1811228 and No.\ PHY-1912632 to RIT; as well as Grants No.\ PHY-1806596, PHY-2110352, and OAC-2004311 to U.\ of Idaho.

We gratefully acknowledge NASA for financial support from Grants No.\ NASA NNH17ZDA001N-TCAN-17-TCAN17-0018 80NSSC18K1488 to RIT, and No. ISFM-80NSSC18K0538 to U.~of Idaho. V.M. is supported by the Exascale Computing Project (17-SC-20-SC), a collaborative effort of the U.S. Department of Energy (DOE) Office of Science and the National Nuclear Security Administration. Work at Oak Ridge National Laboratory is supported under Contract No. DE-AC05-00OR22725 with the U.S. Department of Energy. 
Computational resources were provided by TACC's Frontera supercomputer allocations (Grant No. PHY-20010 and No. AST-20021). Additional resources were provided by RIT's BlueSky and Green Pairie and Lagoon Clusters acquired with NSF Grants No. PHY-2018420, No. PHY-0722703, No. PHY-1229173, and No. PHY-1726215. Funding for computer equipment to support the development of \SENR was provided in part by NSF EPSCoR Grant No. OIA-1458952 to West Virginia University. All plots in this paper were created using {\sc Matplotlib}~\cite{2007CSE.....9...90H} for which we have used the {\sc PyCactus} \cite{2021ascl.soft07017K} to import {\sc Carpet} data.

\end{acknowledgments}

\bibliographystyle{apsrev4-1}
\bibliography{ref}

\end{document}